\documentclass[12pt]{article}
\pdfoutput=1
\usepackage[]{hyperref}
\usepackage{xcolor}
\usepackage[]{putex}
\usepackage[utf8]{inputenc}
\usepackage{cite}
\usepackage{mathrsfs}

\usepackage{amsmath,amsthm,amssymb}
\usepackage{tikz}
\usepackage{tikz-cd}
\usetikzlibrary{decorations.markings}
\usetikzlibrary{calc}
\usepackage{subcaption}
\usepackage{enumerate}
\usepackage{enumitem}

\numberwithin{equation}{section}

\makeatletter
\DeclareRobustCommand*{\bfseries}{%
  \not@math@alphabet\bfseries\mathbf
  \fontseries\bfdefault\selectfont
  \boldmath
}
\makeatother
\hypersetup{
  colorlinks=true,
  linkcolor=blue,
  citecolor=green!60!black,
  urlcolor=cyan!70!black,
  linktoc=all
  }

\input{glyphtounicode}
\begin{document}

\title{Shadow Celestial Amplitude}
\authors{Chi-Ming Chang$^{1,2}$\footnote{\tt cmchang@tsinghua.edu.cn}, Wei Cui$^{2,1}$\footnote{\tt cwei@bimsa.cn}, Wen-Jie Ma,$^{2,1}$\footnote{\tt wenjia.ma@bimsa.cn}\\
Hongfei Shu$^{2,1}$\footnote{\tt shuphy124@gmail.com} $, $ Hao Zou$^{2,1}$\footnote{\tt hzou@bimsa.edu}}
%\institution{Laval}{$^1$D\'epartement de Physique, de G\'enie Physique et d'Optique,\cr\hskip0.06in Universit\'e Laval, Qu\'ebec, QC G1V 0A6, Canada
%}
\institution{}{$^1$Yau Mathematical Sciences Center, Tsinghua University, Beijing 100084, China}
\institution{}{$^2$Yanqi Lake Beijing Institute of Mathematical Sciences and Applications (BIMSA),\cr Huairou District, Beijing 101408, P. R. China}

\abstract{
We study scattering amplitudes in the shadow conformal primary basis, which satisfies the same defining properties as the original conformal primary basis and has many advantages over it. The shadow celestial amplitudes exhibit locality manifestly on the celestial sphere, and behave like correlation functions in conformal field theory under the operator product expansion (OPE) limit. We study the OPE limits for three-point shadow celestial amplitude, and general $2\to n-2$ shadow celestial amplitudes from a large class of Feynman diagrams.
In particular, we compute the conformal block expansion of the $s$-channel four-point shadow celestial amplitude of massless scalars at tree-level, and show that the expansion coefficients factorize as products of OPE coefficients.
}

%\date{July 2022}

\maketitle

{\hypersetup{linkcolor=black}
\tableofcontents
}

%%%%%%%%%%%%%%%%%%%%%%%%%%%%%%%%%%%%%%%%%%%%%%%%%%%%%%%%%%%%%%%%%%%%%%%%

\section{Introduction} \label{sec:1}

Celestial holography is believed to be a concrete realization of holographic principles for quantum gravity in asymptotically flat spacetime (AFS)  \cite{Pasterski:2016qvg, Strominger:2017zoo,Raclariu:2021zjz, Pasterski:2021rjz, Pasterski:2021raf}. It relates scattering amplitudes of a quantum field theory or quantum gravity in a four-dimensional AFS to correlation functions of a celestial conformal field theory (CCFT) on the two-dimensional celestial sphere. The Lorentz symmetry of the four-dimensional AFS is realized as the $SL(2,\mathbb{C})$ conformal symmetry of the celestial sphere. The goal of celestial holography is to study the scattering amplitudes in AFS by using the techniques developed in conformal field theory (CFT). One of the most important achievements in celestial holography is recasting the soft theorems in flat space into Ward identities in two-dimensional CFTs. The currents associated to these Ward identities generate asymptotic symmetries in the four-dimensional spacetime \cite{Weinberg:1965nx, He:2014laa, Kapec:2016jld, Donnay:2018neh, Fan:2019emx, Pate:2019mfs, Adamo:2019ipt, Puhm:2019zbl, Guevara:2019ypd,Stieberger:2018onx, Fotopoulos:2019tpe} and can be re-organized into the $w_{1+\infty}$ algebra \cite{Strominger:2021lvk, Ball:2021tmb, Strominger:2021mtt}.

To manifest the $SL(2,\mathbb{C})$ Lorentz symmetry in the scattering amplitudes, one needs to change the basis of asymptotic states from the standard plane-wave basis to the \textit{conformal primary basis}\cite{Pasterski:2017kqt, Law:2020tsg, Narayanan:2020amh, Iacobacci:2020por}. By definition, the conformal primary basis must satisfy the equations of motion and transform covariantly under $SL(2,\mathbb{C})$.  
The S-matrix elements in the conformal primary basis are referred to as \textit{celestial amplitudes}. The conformal primary basis that is widely used in the literature for massless particles is built from the usual plane-wave basis followed by a Mellin transformation. However, in this basis, the coordinates on the celestial sphere relate directly to the solid angles of the flat space momentum. Thus the corresponding celestial amplitudes are highly constrained by four-dimensional kinematics and do not take the standard form of CFT correlation functions. For example, the four-point celestial amplitudes of massless scalars contain an unfamiliar delta-function $\delta(\chi-\bar{\chi})$ originated from the momentum conservation. This distributional factor forces the celestial amplitude to live on the equator of the celestial sphere.
In addition, depending on the assignments of the incoming and outgoing particles, the celestial amplitudes are only supported in disjoint intervals on the equator. 

%The massless conformal primary basis also leads to subtleties when one studies the OPE limit of massless celestial amplitudes.\footnote{The OPE analysis based on the conformal primary basis can be found in \cite{Pate:2019lpp, Pate:2019mfs, Fotopoulos:2019tpe, Strominger:2021mtt, Jiang:2021csc, Bhardwaj:2022anh}.} For instance, the OPE limit of the celestial amplitudes involving four-massless particles vanishes at leading order since the momentum conservation of three massless particles has no non-trivial solution in Minkowski space.\fixme{Why this is a problem?}
%Due to this problem, the OPE limit in the literature was studied in Klein space \cite{Pate:2019lpp}.\footnote{Other work which studies the celestial amplitudes in Klein space can be found in \cite{Pasterski:2017ylz,Atanasov:2021oyu,Atanasov:2021cje,Pasterski:2022lsl,De:2022gjn,Hu:2022syq}.}  
%Moreover, by the momentum conservation, taking OPE limit of two massless particles does not commute with the Mellin integral. This can be seen from the OPE analysis of celestial amplitudes involving two massless scalars and one massive scalar.
%More detailed discussion on this non-commutativity can be found in Section \ref{sec:shadow}.

Finally, in terms of the massless conformal primary basis, the celestial amplitudes do not have proper conformal block expansion. This can be seen by looking at the  $s$-channel tree-level celestial amplitude of two incoming, two outgoing massless particles and one massive exchange particle. The imaginary part of the corresponding scattering amplitude in the plane-wave basis is factorized into two three-point scattering amplitudes due to the optical theorem. This leads to a factorization in the conformal partial wave expansion of the celestial amplitudes \cite{Lam:2017ofc,Chang:2021wvv}.\footnote{Partial wave expansion of the celestial amplitudes is also studied in\cite{Nandan:2019jas,Law:2020xcf}.} However, since the integration kernel in the conformal partial wave expansion does not have poles located at the right half-plane, one does not get a conformal block expansion by closing the contour. Again, in the literature, the studies of conformal block expansion of celestial amplitudes are limited to the Klein space \cite{Atanasov:2021cje, Hu:2022syq, Banerjee:2022hgc, De:2022gjn} or three dimensional space \cite{Garcia-Sepulveda:2022lga, Jiang:2022hho}.
%\WJM{38 is not in Klein but in 3D.}

%The celestial amplitudes have the similar form as the correlation functions of the CFT \cite{DiFrancesco:1997nk}. 
%We expect that using standard techniques, one can derive all the CFT data from these celestial amplitudes. For example, the spectrum of primary operators can be extracted %from the conformal block decomposition of the celestial amplitudes
%\cite{Lam:2017ofc, Nandan:2019jas, Law:2020xcf, Atanasov:2021cje, Fan:2021isc, Fan:2021pbp, Hu:2022syq, Banerjee:2022hgc, De:2022gjn}
%and the operator product expansion (OPE) coefficients are determined by the tree-point correlation functions .

To fix these issues, we consider a different set of conformal primary wavefunctions for massless particles, which are in a different branch of solutions to the two defining properties of conformal primary wavefunction \cite{Pasterski:2016qvg}, i.e. they satisfy the equations of motion and transform covariantly under $SL(2,{\mathbb C})$.
It turns out that, up to a constant factor, these conformal primary wavefunctions are equivalent to the shadow transformations \cite{Ferrara:1972uq, Osborn:2012vt} of the original conformal primary wave functions.\footnote{The shadows of the conformal wave functions were previously studied in \cite{Pasterski:2017kqt}.} We will refer to this basis as the {\it shadow conformal primary basis}.
Expanding the scattering amplitude in the shadow conformal primary basis, we define the {\it shadow celestial amplitudes}, which can be obtained by performing the shadow transformation of all external operators of the celestial amplitudes.\footnote{Implementing the shadow transformation on one of the external operators of the celestial amplitude was discussed in \cite{Fan:2021isc,Fan:2021pbp,Fan:2022kpp}.} The shadow celestial amplitudes resolve all the abovementioned issues and lead to the standard correlation functions of CFTs. Specifically, the shadow celestial amplitudes of four massless particles no longer have $\delta(\chi-\bar{\chi})$ and are defined on the entire celestial sphere. In addition, the shadow celestial amplitudes have well-behaved OPE limits. For scattering amplitudes with $n$ external massless real scalars, we consider the OPE limit by making the celestial coordinates of the first two incoming particles close to each other. For amplitudes from a large class of Feynman diagrams, we find that the shadow celestial amplitudes factorize as expected as $n$-point correlation functions in a CFT. Using the generalized optical theorem, we also obtain the correct factorization of the imaginary part of shadow celestial amplitudes for any Feynman diagrams. What's more, we compute the 4-point shadow celestial amplitudes involving four external massless scalars and one exchange massive scalar and derive its conformal block expansion in the $12\leftrightarrow34$ channel. We find that the coefficients appearing in the conformal block expansion give the correct OPE coefficients obtained from the corresponding 3-point shadow celestial amplitudes matching the results from the OPE analysis.

This paper is organized as follows. 
In Section \ref{sec:shadow}, we review the celestial amplitudes, discuss their disadvantages, and introduce the shadow celestial amplitudes.
In Section \ref{sec:OPE}, we analyze the OPE limits of $n$-point shadow celestial amplitudes involving $n$ external real massless scalars and show that the OPE limits are well-defined for the shadow celestial amplitudes. 
In Section \ref{sec:5}, we work out the conformal block expansion on the 4-point shadow celestial amplitudes involving four external massless scalars and one exchange massive scalar. We find the complete agreement between the block expansion coefficients and OPE coefficients. 
In Section \ref{Sec:outlook}, we conclude our work and point out a few future directions. In Appendix \ref{sec:pre}, we review the generalized optical theorem, the shadow transformation and conformal partial waves, that will be used in later parts of this paper.

%Similar ideas 
%Attempts on performing shadow transformation or, for spacetime with $(2,2)$ signature, light ray transformations to celestial amplitudes before the conformal block %expansion have been explored in \cite{Fan:2021isc, Fan:2021pbp} and \cite{Sharma:2021gcz, Hu:2022syq, Banerjee:2022hgc, De:2022gjn}.

\section{Shadow conformal primary basis} \label{sec:shadow}

\subsection{Review on the celestial amplitudes}

Celestial amplitudes are obtained by expanding the position space amplitudes with respect to the conformal primary wavefunctions \cite{Pasterski:2017kqt} instead of the plane-waves, \textit{i.e.,}\footnote{Throughout this paper, $\mathcal{O}_i(z_i)$ should be understood as $\mathcal{O}_i(z_i,\bar{z}_i)$. We use this abbreviation to simplify the notation.}
\begin{align}\label{eq:OCA}
\mathcal{A}^{\Delta_i}(z_i)=\bigg(\prod_{j=1}^{k+n}\int d^{4}x_j\bigg)\bigg(\prod_{j=1}^k\phi_{\Delta_j}^{+}(z_j;x_j)\bigg)\bigg(\prod_{j=k+1}^{k+n}\phi^{-}_{\Delta_j}(z_j;x_{j})\bigg)\mathcal{M}(x_i)\;,
\end{align}
where $\mathcal{M}(x_j)$ is the scattering amplitude in position space. The conformal primary wave functions $\phi^{\pm}_{\Delta}(z;x)$ for massless and massive scalars with mass $m$ are given by
\begin{align}\label{eq:MasslessIn}
\phi^{\pm}_{\Delta}(z;x)= \int_{0}^{+\infty}d\omega\,\omega^{\Delta-1}e^{\pm i\omega\hat{q}\cdot x-\epsilon\omega}\;,
\end{align}
and
\begin{align}\label{eq:MassiveIn}
\phi^{\pm}_{\Delta}(z;x)=\int\frac{d^3\hat{p}^{\prime}}{\hat{p}^{\prime0}}G_{\Delta}(z,\bar{z};\hat{p}^{\prime})e^{\pm im\hat{p}^{\prime}\cdot x}\;,
\end{align}
respectively. Here $z$ and $\bar{z}$ are coordinates on celestial sphere and $G_{\Delta}(z,\bar{z};\hat{p})$ is the bulk-to-boundary propagator. The coordinates $(z,\bar z)$ on the celestial sphere are related to the massless on-shell momenta $q^\mu$ through
\begin{align}\label{hq}
q^{\mu}=\omega\hat{q}^{\mu}=\omega(1+z\bar{z},z+\bar{z},-i(z-\bar{z}),1-z\bar{z})
\end{align}
and to the massive on-shell momenta $p^{\mu}$ through
\begin{align}\label{hp}
p^{\mu}=m\hat{p}^{\mu}=\frac{m}{2y}(1+y^2+z\bar{z},z+\bar{z},-i(z-\bar{z}),1-y^2-z\bar{z})\,.
\end{align}
In terms of $\hat{p}$ in \eqref{hp}, the bulk-to-boundary propagator takes the form \footnote{In this paper, we use the most positive metric in four-dimensional flat space, \textit{i.e.}, $g_{AB}=\text{diag}(-1,+1,+1,+1)$.}
\begin{align}\label{eqn:bdry_to_bulk}
G_{\Delta}(\hat{q};\hat{p}^{\prime})=\frac{1}{(-\hat{q}\cdot\hat{p}^{\prime})^{\Delta}}\;.
\end{align}
In terms of the conformal primary basis \eqref{eq:MasslessIn}, the celestial amplitudes in \eqref{eq:OCA} with $n_1$ massless scalars and $n_2$ massive scalars can be re-expressed as
\begin{align}\label{eq:OCA1}
\mathcal{A}^{\Delta_i}(z_i)=\bigg(\prod_{j=1}^{n_1}\int d\omega_j\omega_j^{\Delta_j-1}\bigg)\bigg(\prod_{j=1}^{n_2}\int\frac{d^3\hat{p}^{\prime}_j}{\hat{p}^{\prime0}_j}G_{\Delta_j}(\hat{q}_j;\hat{p}^{\prime}_j)\bigg)\mathcal{M}_{k\rightarrow n}(q_i,p^{\prime}_i)\;,
\end{align}
where $\mathcal{M}(q_i,p^{\prime}_i)$ is the scattering amplitude in the plane-wave basis.

\subsection{Disadvantages of the celestial amplitudes}

Unfortunately, the celestial amplitudes defined in \eqref{eq:OCA1} have many disadvantages. For example, the four-point celestial amplitudes $\mathcal{A}^{\Delta_i}_{12\rightarrow34}$ that describe the two-to-two scattering of four massless particles are exotic. The form of $\mathcal{A}^{\Delta_i}_{12\rightarrow34}$ is completely fixed by conformal symmetry up to a function of conformal cross-ratios. In other words, we have\footnote{Again, we omit the anti-holomorphic dependence to simplify the notation.}
\begin{align}\label{eq:A}
    \mathcal{A}^{\Delta_i}_{12\rightarrow34}(z_{i})=I_{12-34}(z_{i})f(\chi)\;,
\end{align}
where $f(\chi)$ is a function of cross-ratios
\begin{align}
\chi=\frac{z_{12}z_{34}}{z_{13}z_{24}}\;,\hspace{0.5cm}\bar{\chi}=\frac{\bar{z}_{12}\bar{z}_{34}}{\bar{z}_{13}\bar{z}_{24}}\;,
\end{align}
and 
$I_{12-34}$ is given by
\begin{align}\label{eq:I1234}
  I_{12-34}\equiv\dfrac{\left(\dfrac{z_{24}}{z_{14}} \right)^{h_{12}} \left(\dfrac{z_{14}}{z_{13}} \right)^{h_{34}}}{z_{12}^{h_1 + h_2} z_{34}^{h_3+h_4}}\dfrac{\left(\dfrac{\bar{z}_{24}}{\bar{z}_{14}} \right)^{\bar{h}_{12}} \left(\dfrac{\bar{z}_{14}}{\bar{z}_{13}} \right)^{\bar{h}_{34}}}{\bar{z}_{12}^{\bar{h}_1+\bar{h}_2} \bar{z}_{34}^{\bar{h}_3+\bar{h}_4}}\;,\quad h_{ij}\equiv h_i-h_j\;,\quad\bar{h}_{ij}\equiv \bar{h}_i-\bar{h}_j\;,
\end{align}
and $\Delta_i=2h_i$.
The corresponding scattering amplitude in the plane-wave basis at tree level is given by the diagrams in $s$-, $t$-, and $u$-channels as
\begin{align}
\mathcal{M}_{2\to2}(q'_i)=(2\pi)^4\delta^{(4)}(q^{\prime}_1+q^{\prime}_2-q^{\prime}_3-q^{\prime}_4)(-ig)^2\left(\frac{1}{s-m^2}+\frac{1}{t-m^2}+\frac{1}{u-m^2}\right)\;,
\end{align}
where $m$ is the mass of the exchange massive scalar.
This leads to the following expression for the function $f(\chi)$ with $\chi\geq1$
\begin{align}\label{eq:4pt}
    f(\chi)=N_{gm}(\beta)\chi^{2}(\chi-1)^{h_{13}-h_{24}}\delta(\chi-\bar{\chi})\bigg[1+e^{-i\pi\frac{\beta}{2}}\bigg(\frac{\chi}{\chi-1}\bigg)^{\frac{\beta}{2}}+e^{-i\pi\frac{\beta}{2}}\chi^{\frac{\beta}{2}}\bigg]\;,
\end{align}
where $h_{ij}\equiv h_i-h_j$ and $\beta=\sum_{i=1}^4\Delta_i-4$ and $N_{gm}(\beta)$ is given by
\begin{align}
N_{gm}(\beta)= \frac{g^2\pi^5m^{\beta-2}}{2^{\beta-1}}\frac{e^{i\pi\beta/2}}{\sin \pi\beta/2}\;.
\end{align}

From \eqref{eq:4pt}, we note that $\mathcal{A}^{\Delta_i}_{12\rightarrow34}(z_{i})$ contain a delta-function $\delta(\chi-\bar{\chi})$ which forces the four-point celestial amplitudes to live on the equator of the celestial sphere and makes the structure of correlation functions in CCFTs very different from those in the standard CFTs. Moreover, \eqref{eq:4pt} is for the $12\to34$ kinematic and only valid when $\chi\geq 1$. The other two kinematics $14\to23$ and $13\to24$ are defined in distinct intervals $\chi\le0$ and $0\le\chi\le1$ on the equator. This also makes the four-point celestial amplitudes exotic. Finally, $\mathcal{A}^{\Delta_i}_{12\rightarrow34}(z_{i})$ does not have a proper conformal block expansion. We take the $s$-channel scattering amplitude (the first term in \eqref{eq:4pt}) as an example. As we will see in Section \ref{sec:5}, the partial wave expansion of the $s$-channel celestial amplitudes $\mathcal{A}^{\Delta_i}_{s}$ takes the form\footnote{See Appendix \ref{sec:PW} for a review on the conformal block and conformal partial wave.}
\begin{align}\label{eq:CPE4ptA}
 \mathcal{A}^{\Delta_i}_{s}(z_{i})=\int_{-\infty}^{+\infty}\frac{d\lambda}{2\pi}\frac{\rho^{\Delta_i}_{h,\bar{h}}}{n_{h,\bar{h}}}K^{\Delta_3,\Delta_4}_{1-h,1-\bar{h}}G^{\Delta_i}_{h,\bar{h}}(z_i)\;,
\end{align}
where $h=\bar{h}=(1+i\lambda)/2$ and the spectral density is 
\begin{align}\label{eq:SD4pt}
\begin{split}
K^{\Delta_3,\Delta_4}_{1-h,1-\bar{h}}\frac{\rho_{h,\bar{h}}^{\Delta_i}}{n_{h,\bar{h}}}=& \frac{-g^2\pi^9m^{\beta-4}}{2^{\beta-6}}\frac{e^{i\pi\beta/2}}{\sin \pi\beta/2}\\
&\times B(h+h_{12},h-h_{12})B(h-h_{34},h+h_{34})D^{\Delta\Delta}(m)\;.
\end{split}
\end{align}
When $\chi\leq 1$, closing the contour to the right-half $\lambda$-plane leads to a vanishing conformal block expansion because the integrand in \eqref{eq:CPE4ptA} does not have poles located in the right-half $\lambda$-plane. On the other hand, when $\chi>1$ the integrand in \eqref{eq:CPE4ptA} does not decay to zero when $\text{Re}(\lambda)$ approaches infinity and we can not close the contour. This reflects the fact that $\mathcal{A}^{\Delta_i}_{s}(z_{i})$ only supports on $\chi>1$. Thus, we conclude that $\mathcal{A}^{\Delta_i}_{s}(z_{i})$ does not have a proper conformal block expansion.

%Besides the above unexpected features of four-point celestial amplitudes, the OPE limit of two massless scalars also are subtle due to the existence of  momentum conservation. One example is the three-point celestial amplitude with two incoming massless scalars and one outgoing massive scalar. This three-point celestial amplitude at tree level takes the form as \textcolor{blue}{[? reference here]}
%
%\begin{align}\label{eq:OCA3pt}
%\mathcal{A}_{2\rightarrow 1}^{\Delta_i}(z_i)=-g\frac{\pi^4m^{\Delta_1+\Delta_2-4}}{2^{\Delta_1+\Delta_2-5}}\frac{B(\frac{\Delta_1+\Delta_{12}}{2},\frac{\Delta_1-\Delta_{12}}{2})}{|z_{12}|^{\Delta_1+\%Delta_2-\Delta_3}|z_{13}|^{\Delta_1-\Delta_2+\Delta_3}|z_{23}|^{\Delta_{2}+\Delta_3-\Delta_1}}\;,
%\end{align}
%
%where $B(a,b)\equiv\Gamma[a]\Gamma[b]/\Gamma[a+b]$ is the beta-function. Taking OPE limit $(z_1,\bar z_1)\rightarrow (z_2,\bar z_2)$ of \eqref{eq:OCA3pt} leads to a standard two-point function. However, taking the same OPE limit in \eqref{eq:OCA1} vanishes due to momentum conservation.\fixme{Spell this out a bit?} This indicates that the integral in \eqref{eq:OCA1} does not commute with the OPE limit. Indeed, one must be very careful to take OPE limit in \eqref{eq:OCA1} since the OPE limit in celestial coordinates is equivalent to colinear limit in momentum. The latter is strongly constrained by on-shell condition and momentum conservation. 

\subsection{A different conformal primary basis}

It is easy to see that the above disadvantages of the celestial amplitudes are due to the fact that the conformal primary wave functions \eqref{eq:MasslessIn} for massless scalars are constructed from the plane-waves by performing integral only over the energy $\omega$. Thus, we consider a different set of conformal primary wave functions for massless scalars:
\begin{align}\label{eq:NewMassless}
\widetilde\phi^{\pm}_{\Delta}(z;x)= \int\frac{d^3q^{\prime}}{q^{\prime0}}G_{\Delta}(\hat{q};q^{\prime})e^{\pm i\omega\hat{q}^{\prime}\cdot x-\epsilon\omega}\;.
\end{align}
Here, $q^\mu$ is defined in \eqref{hq} and $G_{\Delta}(\hat{q};q^{\prime})$ is given by
\begin{align}
 G_{\Delta}(\hat{q};q^{\prime})=\frac{1}{(-\hat{q}\cdot q^{\prime})^{\Delta}}\;,
\end{align}
which takes the same form as bulk-to-boundary propagator \eqref{eqn:bdry_to_bulk} with $\hat{p}^{\prime}$ replaced by $q^{\prime}$. Using the identity $q^{\prime}(z^{\prime})\cdot\hat{q}(z)=-2\omega|z^{\prime}-z|^2$ and noting that
\begin{align}\label{eq:q-z}
\int\frac{d^{3}q}{q^0}=4\int d^2z\int_{0}^{+\infty}d\omega\,\omega\;,
\end{align}
the conformal primary waves functions \eqref{eq:NewMassless} can be written as
\begin{align}\label{eq:NewMassless_as_shadow}
\widetilde\phi^{\pm}_{\Delta}(z;x)= \int d^2z^{\prime}\frac{2^{1-\Delta}}{|z-z^{\prime}|^{2\Delta}}\int_0^{\infty}d\omega\,\omega^{1-\Delta}e^{\pm i\omega\hat{q}^{\prime}\cdot x-\epsilon\omega}=\int d^2z^{\prime}\frac{2^{1-\Delta}}{|z-z^{\prime}|^{2\Delta}}\phi^{\pm}_{\Delta}(z';x)\;,
\end{align}
which are proportional to the shadows of the original conformal primary wave functions $\phi^{\pm}_{\Delta}(z;x)$, that were previously studied in \cite{Pasterski:2017kqt}.
We dub it \textit{the shadow conformal primary basis}.
The shadow conformal primary wave functions \eqref{eq:NewMassless} obey the massless Klein-Gordon equation and transform covariantly under the conformal transformation since the shadow transformation is conformally covariant. 
In other words, they satisfy the two defining properties for conformal primary wave functions given in \cite{Pasterski:2016qvg,Pasterski:2017kqt}.
On the other hand, one does not obtain a new wave function by performing the shadow transformation on the massive conformal primary basis \eqref{eq:MassiveIn} because the result from the shadow integral takes the same form as \eqref{eq:MassiveIn} up to a change of conformal dimension from $\Delta$ to $2-\Delta$ \cite{Pasterski:2017kqt}. 

Using the shadow conformal primary wave functions \eqref{eq:NewMassless}, we define \textit{the shadow celestial amplitudes} as
\begin{align}\label{eq:NCA}
\widetilde{\mathcal{A}}^{\Delta_i}(z_i)=\bigg(\prod_{j=1}^{n}\int\frac{d^3k^{\prime}_j}{k^{\prime0}_j}G_{\Delta_j}(\hat{q}_j;k^{\prime}_j)\bigg)\mathcal{M}(k^{\prime}_i)\;,
\end{align}
where $k^{\prime}_j$ can be either $q^{\prime}_j$ or $\hat{p}^{\prime}_j$ depending on whether the particle is massless or massive. 

\subsection{Translation symmetry}

Unlike the scattering amplitudes in the plane-wave basis, translation symmetry in shadow celestial amplitudes $\widetilde{\mathcal{A}}^{\Delta_i}(z_i)$ becomes obscure, while Lorentz symmetry is manifest. In this subsection, we will discuss how the translation generators $\hat{P}^{\mu}$ act on the shadow celestial amplitude. The action of the translation generators on massive scalars was studied in \cite{Law:2019glh}. Specifically, $\hat{P}^{\mu}$ act on a massive scalar in CCFT as a differential operator $\mathcal{P}^{\mu}$
\begin{align}\label{eq:TMassive}
[\hat{P}^{\mu},\Phi_a^{\Delta_a}(z_a)]=\mathcal{P}_a^{\mu}\Phi_a^{\Delta_a}(z_a)\;,    
\end{align}
where the different operator $\mathcal{P}_a^{\mu}$ acting on massive scalars takes the form
\begin{align}\label{eq:trans-massive}
\mathcal{P}^{\mu}_a=\epsilon_a\frac{m_a}{2}\bigg[\bigg((\partial_{z_a}\partial_{\bar{z}_a}\hat{q}_a^{\mu})+\frac{(\partial_{z_a}\hat{q}_a^{\mu})\partial_{\bar{z}_a}+(\partial_{\bar{z}_a}\hat{q}_a^{\mu})\partial_{z_a}}{\Delta_a-1}+\frac{\hat{q}_a^{\mu}\partial_{z_a}\partial_{\bar{z}_a}}{(\Delta_a-1)^2}\bigg)e^{-\partial_{\Delta_a}}+\frac{\Delta_a\hat{q}_a^{\mu}}{\Delta_a-1}e^{\partial_{\Delta_a}}\bigg]
\end{align}
with $\epsilon_k=\pm 1$ labeling incoming and outgoing states, respectively. 

To find the action of $\hat{P}^{\mu}$ on massless shadow celestial amplitudes \eqref{eq:NCA}, we start with the action in the plane-wave basis
\begin{align}\label{eq:PM}
\hat{P}^{\mu}_a\mathcal{M}(q_a)=\epsilon_a q^{\mu}_a\mathcal{M}(q_a)\;,
\end{align}
Substituting \eqref{eq:PM} into \eqref{eq:NCA} leads to
\begin{align}\label{eq:TM1}
\hat{P}_a^{\mu}\widetilde{\mathcal{A}}^{\Delta_i}(z_i)=\bigg(\prod_{j\neq a}^{n}\int\frac{d^3k^{\prime}_j}{k^{\prime0}_j}G_{\Delta_j}(\hat{q}_j;k^{\prime}_j)\bigg)\int\frac{d^3q_a^{\prime}}{q^{\prime0}_a}\frac{1}{(-\hat{q}_a\cdot q^{\prime}_a)^{\Delta_a}}\epsilon_aq^{\prime\mu}_a\mathcal{M}(q^{\prime}_a)\;.
\end{align}
Thus, finding the action of $\hat{P}^{\mu}$ on massless shadow celestial amplitudes is translated into finding a differential operator $\mathcal{P}^{\mu}_a$ such that
\begin{align}
\mathcal{P}^{\mu}_a\frac{1}{(-\hat{q}_a\cdot q^{\prime}_a)^{\Delta_a}}=\epsilon_aq^{\prime\mu}_a\frac{1}{(-\hat{q}_a\cdot q^{\prime}_a)^{\Delta_a}}\;.   
\end{align}
This can be realized by the following differential operator
\begin{align}\label{eq:TMassless}
\mathcal{P}_a^{\mu}=\frac{1}{2}\epsilon_a\bigg((\partial_{z_a}\partial_{\bar{z}_a}\hat{q}_a^{\mu})+\frac{(\partial_{z_a}\hat{q}^{\mu}_a)\partial_{\bar{z}_a}+(\partial_{\bar{z}_a}\hat{q}^{\mu}_a)\partial_{z_a}}{\Delta_a-1}+\frac{\hat{q}^{\mu}_a\partial_{z_a}\partial_{\bar{z}_a}}{(\Delta_a-1)^2}\bigg)e^{-\partial_{\Delta_a}}\;,
\end{align}
which coincides with the first term of the massive translation operator \eqref{eq:trans-massive} up to the mass $m_a$.
In other words, the action of the translation generators $\hat{P}^{\mu}$ on a massless scalar operator $\mathcal{O}_a^{\Delta_a}(z_a)$ in CCFT is 
\begin{align}
[\hat{P}^{\mu},\mathcal{O}_a^{\Delta_a}(z_a)]=\mathcal{P}_a^{\mu}\mathcal{O}_a^{\Delta_a}(z_a)\;.  
\end{align}
Thus, the translation symmetries amount to the following differential equations 
\begin{align}\label{eq:TS}
\sum_{a=1}^n\mathcal{P}^{\mu}_a\widetilde{\mathcal{A}}^{\Delta_i}_n(z_i)=0\;,    
\end{align}
where $\mathcal{P}_a^{\mu}$ is given in \eqref{eq:trans-massive} when it acts on massive scalar operators and in \eqref{eq:TMassless} when it acts on massless scalar operators.

\section{OPE behaviour of the shadow celestial amplitudes}\label{sec:OPE}

In this section, we will study the OPE behaviour of the shadow celestial amplitudes of $2$ incoming massless real scalars, labeled by $1$ and $2$, and $n-2$ outgoing massless real scalars. Similar analysis works for the same shadow celestial amplitudes with the incoming and outgoing particles swapped. The shadow celestial amplitudes of interest can be written as\footnote{See Appendix \ref{sec:pre} for the convention of the scattering amplitude in the plane-wave basis.}
\begin{align}
\begin{split}
\widetilde{\mathcal{A}}^{\Delta_i}_{2\rightarrow n-2}(z_i)=&2^{1-\Delta_{2}}\int\frac{d^{3}q^{\prime }_{1}}{q^{\prime0}_1}\frac{1}{(-\hat{q}_{1}\cdot q^{\prime}_1)^{\Delta_{1}}}\int d^2z^{\prime}_{2}\frac{1}{|z^{\prime}_2-z_{2}|^{2\Delta_{2}}}\int_{0}^{+\infty}d\omega_2\,\omega_2^{1-\Delta_2}\\
&\times\bigg(\prod_{k=3}^n\int\frac{d^3q^{\prime }_k}{q^{\prime0}_k}G_{\Delta_k}(\hat{q}_k;q^{\prime}_k)\bigg)\mathcal{M}_{2\rightarrow n-2}(q^{\prime}_i)\;,
\end{split}
\end{align}
where we have used \eqref{eq:q-z}. After changing of variables, we have
\begin{align}\label{eq:OPE1}
\begin{split}
\widetilde{\mathcal{A}}^{\Delta_i}_{2\rightarrow n-2}(z_i)&=2^{\Delta_1+\Delta_{2}}\int d^{4}p\int D^2\hat{q}^{\prime}_2\frac{|p|^{2-2\Delta_1-2\Delta_2}}{(-\hat{q}_2\cdot\hat{q}^{\prime}_2)^{\Delta_2}(-\hat{q}^{\prime}_2\cdot p)^{2-\Delta_1-\Delta_2}(-\hat{q}^{\prime}_2\cdot Y)^{\Delta_1}}\\
&\times\bigg(\prod_{k=3}^n\int\frac{d^3q^{\prime }_k}{q^{\prime0}_k}G_{\Delta_k}(\hat{q}_k;q^{\prime}_k)\bigg)\mathcal{M}_{2\rightarrow n-2}(q^{\prime}_i)\;,
\end{split}
\end{align}
where $q^{\prime}_1+q^{\prime}_2\equiv p\equiv M\hat{p}$ with $\hat{p}\cdot\hat{p}=-1$, $D^2\hat{q}^{\prime}\equiv d^2z^{\prime}$ and $Y^{\mu}\equiv2(-\hat{q}_{1}\cdot\hat{p})\hat{p}^{\mu}-\hat{q}_{1}^{\mu}$, and note that $\omega_2=\frac{p^2}{2\hat{q}^{\prime}_2\cdot p}$. We have also used the relation
\begin{align}
\int\frac{d^{3}q^{\prime }_{1}}{q^{\prime0}_1}\int d^2z^{\prime}_{2}\frac{1}{|z^{\prime}_2-z_{2}|^{2\Delta_{2}}}\int_{0}^{+\infty}d\omega_2=2^{\Delta_2}\int d^{4}p\int\frac{D^2\hat{q}^{\prime}_2}{-\hat{q}_2^\prime \cdot p}\frac{1}{(-\hat{q}_2\cdot\hat{q}^{\prime}_2)^{\Delta_2}}\;.
\end{align}

In the following subsections, we will start with \eqref{eq:OPE1} and study the OPE behaviours for different scattering amplitudes $\mathcal{M}_{2\rightarrow n-2}(q^{\prime}_1,q^{\prime}_2,\cdots,q^{\prime}_n)$. 

\subsection{OPE analysis for three-point shadow celestial amplitudes}\label{sec:OPE_3-point}

We first focus on the three-point shadow celestial amplitudes. Since scattering amplitudes involving three massless scalars cannot satisfy momentum conservation, we will consider the shadow celestial amplitudes for two incoming massless scalars and one outgoing massive scalar. In this case, the scattering amplitudes are
\begin{align}
i\mathcal{M}_{2\rightarrow 1}(q^{\prime}_1,q^{\prime}_2,p^{\prime}_3)=i(2\pi)^4\delta^{(4)}(p-p^{\prime}_3)\mathcal{T}_{2\rightarrow 1}(p^2)\;,
\end{align}
where we defined $p=q^{\prime}_1+q^{\prime}_2$. Plugging the above express into \eqref{eq:OPE1} and changing integral variables from $p^{\mu}$ to $M\hat{p}^{\mu}$ then lead to
\begin{align}
\begin{split}
\widetilde{\mathcal{A}}^{\Delta_i}_{2\rightarrow 1}(z_i)=&2^{\Delta_{1}+\Delta_{2}}\int_0^{+\infty}dM\int\frac{d^{3}\hat{p}}{\hat{p}^0}\int D^2\hat{q}^{\prime}_2\frac{M^{3-\Delta_1-\Delta_2}}{(-\hat{q}_2\cdot\hat{q}^{\prime}_2)^{\Delta_2}(-\hat{q}^{\prime}_2\cdot\hat{p})^{2-\Delta_1-\Delta_2}(-\hat{q}^{\prime}_2\cdot Y)^{\Delta_1}}\\
&\qquad\times\int\frac{d^3\hat{p}^{\prime}_3}{\hat{p}^{\prime0}_3}G_{\Delta_3}(\hat{q}_3;\hat{p}^{\prime}_3)(2\pi)^4\delta^{(4)}(p-p^{\prime}_3)\mathcal{T}_{2\rightarrow 1}(M^2)\;.
\end{split}
\end{align}
In Appendix \ref{sec:ConformalIntegral}, we compute the integral over $\hat{q}^{\prime}_2$ and expand the results around small $\hat{q}_{12}\equiv - \frac{1}{2}\hat{q}_1\cdot\hat{q}_2 $. Substituting the expansion \eqref{eq:ConformalI}, we get
\begin{align}
\begin{split}
\widetilde{\mathcal{A}}^{\Delta_i}_{2\rightarrow 1}=&\frac{(2\pi)^5\Gamma[1-\Delta_1]\Gamma[1-\Delta_2]}{\Gamma[2-\Delta_1-\Delta_2]}\sum_{n=0}^{\infty}\frac{(\Delta_1)_n(\Delta_2)_n}{(n!)^2}\int_0^{+\infty}dMM^{3-\Delta_1-\Delta_2}\mathcal{T}_{2\rightarrow 1}(M^2)\\
&\times (\hat{q}_{12})^n\int\frac{d^{3}\hat{p}}{\hat{p}^0}\int\frac{d^3\hat{p}^{\prime}_3}{\hat{p}^{\prime0}_3}G_{\Delta_1+n}(\hat{q}_1;\hat{p})G_{\Delta_2+n}(\hat{q}_2;\hat{p})G_{\Delta_3}(\hat{q}_3;\hat{p}^{\prime}_3)\delta^{(4)}(p-p^{\prime}_3)\;.
\end{split}
\end{align}
Rewriting the delta-function as
\begin{align}
\delta^{(4)}(p-p^{\prime}_3)=M^{-3}\hat{p}^0\delta(M-m)\delta^{(3)}(\hat{p}-\hat{p}^{\prime}_3) = M^{-1}{p}^0\delta(M-m)\delta^{(3)}({p}-{p}^{\prime}_3)
\end{align}
and performing the integral over $M$ and $\hat{p}^{\prime}_3$ lead to
\begin{align}\label{eq:OPE3pt}
\begin{split}
\widetilde{\mathcal{A}}^{\Delta_i}_{2\rightarrow 1}(z_i)=&\frac{(2\pi)^5\Gamma[1-\Delta_1]\Gamma[1-\Delta_2]m^{-\Delta_1-\Delta_2}\mathcal{T}_{2\rightarrow 1}(m^2)}{\Gamma[2-\Delta_1-\Delta_2]}\sum_{n=0}^{\infty}\frac{(\Delta_1)_n(\Delta_2)_n}{(n!)^2}(\hat{q}_{12})^n\\
&\times\int\frac{d^{3}\hat{p}}{\hat{p}^0}G_{\Delta_1+n}(\hat{q}_1;\hat{p})G_{\Delta_2+n}(\hat{q}_2;\hat{p})G_{\Delta_3}(\hat{q}_3;\hat{p})\;.
\end{split}
\end{align}
One can immediately recognize that the second line is exactly the three-point AdS Witten diagram $W_{\Delta_1+n,\Delta_2+n,\Delta_3}(\hat{q}_i)$. In the short distance limit $\hat{q}_1\rightarrow\hat{q}_2$, $W_{\Delta_1+n,\Delta_2+n,\Delta_3}(\hat{q}_i)$ behaves as %\footnote{We omit a term proportional to $\delta(\Delta_1+\Delta_2-2)\delta^{(2)}(z_1-z_2)$. This term can be removed by setting $\Delta_1+\Delta_2\neq 2$ }
\begin{align}
\begin{split}
W_{\Delta_1+n,\Delta_2+n,\Delta_3}(\hat{q}_i)=&\frac{\Gamma[\frac{\Delta_1+\Delta_{23}}{2}+n]\Gamma[\frac{\Delta_{12}+\Delta_3}{2}]\Gamma[\frac{-\Delta_{12}+\Delta_3}{2}]\Gamma[\frac{\sum_{i=1}^3\Delta_i-2}{2}+n]}{\Gamma[\Delta_1+n]\Gamma[\Delta_2+n]\Gamma[\Delta_3]}\\
&\qquad\times(\hat{q}_{12})^{\frac{-\Delta_1-\Delta_2+\Delta_3}{2}-n}(\hat{q}_{23})^{-\Delta_3}\left[1+O(\hat{q}_{12})\right]\;.
\end{split}
\end{align}
We arrive
\begin{align}\label{eq:OPE3pt}
\begin{split}
\widetilde{\mathcal{A}}^{\Delta_i}_{2\rightarrow 1}(z_i)=&\frac{(2\pi)^5\Gamma[1-\Delta_1]\Gamma[1-\Delta_2]m^{-\Delta_1-\Delta_2}\mathcal{T}_{2\rightarrow 1}(m^2)}{\Gamma[2-\Delta_1-\Delta_2]}(\hat{q}_{12})^{\frac{-\Delta_1-\Delta_2+\Delta_3}{2}}(\hat{q}_{23})^{-\Delta_3}\\
&\times\sum_{n=0}^{\infty}\frac{\Gamma[\frac{\Delta_1+\Delta_{23}}{2}+n]\Gamma[\frac{\Delta_{12}+\Delta_3}{2}]\Gamma[\frac{-\Delta_{12}+\Delta_3}{2}]\Gamma[\frac{\sum_{i=1}^3\Delta_i-2}{2}+n]}{(n!)^2\Gamma[\Delta_1]\Gamma[\Delta_2]\Gamma[\Delta_3]}[1+O(\hat{q}_{12})]\;.
\end{split}
\end{align}
Finally, computing the summation over $n$ gives
\begin{align}\label{eq:Ex3pt}
\begin{split}
\widetilde{\mathcal{A}}^{\Delta_i}_{2\rightarrow 1}=&C_{\Delta_1,\Delta_2,\Delta_3}(m)\hat{q}_{12}^{\frac{-\Delta_1-\Delta_2+\Delta_3}{2}}\hat{q}_{23}^{-\Delta_3}[1+O(\hat{q}_{12})]\;,
\end{split}
\end{align}
where the coefficient $C_{\Delta_1,\Delta_2,\Delta_3}(m)$ is
\begin{align}\label{eq:C3pt1}
C_{\Delta_1,\Delta_2,\Delta_3}(m)=\frac{\mathcal{T}_{2\rightarrow 1}(m^2)\Gamma[1-\Delta_1]\Gamma[1-\Delta_2]\Gamma[\frac{\Delta_1+\Delta_{23}}{2}]\Gamma[\frac{\Delta_{12}+\Delta_3}{2}]\Gamma[\frac{\Delta_3-\Delta_{12}}{2}]\Gamma[\frac{\sum_{i=1}^3\Delta_i-2}{2}]}{(2\pi)^{-5}m^{\Delta_1+\Delta_2}\Gamma[\Delta_1]\Gamma[\Delta_2]\Gamma[\Delta_3]\Gamma[\frac{2-\Delta_1-\Delta_{23}}{2}]\Gamma[\frac{4-\sum_{i=1}^3\Delta_i}{2}]}
\end{align}
with $\Delta_{ij}\equiv\Delta_i-\Delta_j$. From \eqref{eq:Ex3pt}, we find that in the OPE limit $\hat{q}_1\rightarrow\hat{q}_2$, $\widetilde{\mathcal{A}}^{\Delta_i}_{2\rightarrow 1}$ behave as
\begin{align}\label{eq:OPE3pt}
\lim_{\hat{q}_1\rightarrow\hat{q}_2}\bigg(\hat{q}_{12}^{\frac{\Delta_1+\Delta_2-\Delta_3}{2}}\widetilde{\mathcal{A}}^{\Delta_i}_{2\rightarrow 1}(z_i)\bigg)=C_{\Delta_1,\Delta_2,\Delta_3}(m)\hat{q}_{23}^{-\Delta_3}.
\end{align}
However, since the OPE coefficient \eqref{eq:C3pt1} diverges when $\Delta_1+\Delta_2-\Delta_3=-2n$ with $n\in\mathbb{Z}_{\geq0}$, the above limit is only well-defined when $\Delta_1+\Delta_2-\Delta_3\neq -2n$. 

When $\Delta_1+\Delta_2-\Delta_3=-2n+2i\nu$ with $\nu\in\mathbb{R}$, we use the following formula
\begin{align}
  \lim_{\hat{q}_1\rightarrow\hat{q}_2}\bigg(\Gamma[-n+i\nu]\hat{q}_{12}^{-i\nu}\bigg)=\frac{\Gamma[-n+i\nu]}{\Gamma[i\nu]}\lim_{\hat{q}_1\rightarrow\hat{q}_2}\int_{0}^{\infty}dx\;x^{i\nu-1}e^{-x\hat{q}_{12}}=\frac{-2\pi(-1)^n}{n!}\delta(\nu)\;.  
\end{align}
This leads to
\begin{align}\label{eq:OPE3pt1}
\lim_{\hat{q}_1\rightarrow\hat{q}_2}\bigg(\hat{q}_{12}^{-n}\widetilde{\mathcal{A}}^{\Delta_1,\Delta_2,\Delta_1+\Delta_2+2n+i\nu}_{2\rightarrow 1}(z_i)\bigg)=-2\pi\underset{\Delta_3=\Delta_{1}+\Delta_2+2n}{\text{Res}}\bigg(C_{\Delta_1,\Delta_2,\Delta_3}(m)\bigg)\delta(\nu)\hat{q}_{23}^{-\Delta_3}\;.
\end{align}
For later convenience, we define the OPE coefficient $\mathcal{C}_{\Delta_1,\Delta_2,\Delta_1+\Delta_2+2n}(m)$ as
\begin{align}\label{eq:C3pt}
 \mathcal{C}_{\Delta_1,\Delta_2,\Delta_3}(m)=
 \begin{cases}
 -2\pi\underset{\Delta_3=\Delta_{1}+\Delta_2+2n}{\text{Res}}\bigg(C_{\Delta_1,\Delta_2,\Delta_3}(m)\bigg)&\text{for}~\Delta_3=\Delta_1+\Delta_2+2n,\\
 C_{\Delta_1,\Delta_2,\Delta_3}(m)&\text{for}~\Delta_3\neq\Delta_1+\Delta_2+2n,
 \end{cases}
 \end{align}
with $n\in \mathbb{Z}_{\geq0}$.

From \eqref{eq:OPE3pt1} and \eqref{eq:C3pt} we see that apart from the delta-function $\delta(\nu)$, the OPE limit of $\widetilde{\mathcal{A}}^{\Delta_1,\Delta_2,\Delta_3}_{2\rightarrow 1}(z_i)$ is finite when $\Delta_3=\Delta_1+\Delta_2$. In the following subsections, we will focus on the finite parts in the OPE limit and study the OPE behaviour of $n\geq3$-point shadow celestial amplitudes.

%
%We note that although we have expanded $\widetilde{\mathcal{A}}^{\Delta_i}_{2\rightarrow 1}(z_i)$ around $\hat{q}_{12}$ which led to the summation over $n$, different value of $n$ finally contributes to the same order of $\hat{q}_{12}$ with powers $\frac{-\Delta_1-\Delta_2+\Delta_3}{2}$. Thus in order to get the correct OPE limit in three-point shadow celestial amplitudes, we have to take into account the contribution from all value of $n$. This makes the OPE behaviour of three-point shadow celestial amplitudes a bit different from the higher-point shadow celestial amplitudes. 

\subsection{OPE analysis from the generalized optical theorem}\label{ssec:OPEO}

The leading order OPE behaviour can be obtained through replacing all $\hat{q}_1$ in \eqref{eq:OPE1} by $\hat{q}_2$, leading to
\begin{align}\label{eq:OPE3}
\begin{split}
\lim_{\hat{q}_1\rightarrow\hat{q}_2}\widetilde{\mathcal{A}}^{\Delta_i}_{2\rightarrow n-2}(z_i)&=2^{\Delta_1+\Delta_{2}}\int d^{4}p\int D^2\hat{q}^{\prime}_2\frac{|p|^{2-2\Delta_1-2\Delta_2}}{(-\hat{q}_2\cdot\hat{q}^{\prime}_2)^{\Delta_2}(-\hat{q}^{\prime}_2\cdot p)^{2-\Delta_1-\Delta_2}(-\hat{q}^{\prime}_2\cdot Y^{\prime})^{\Delta_1}}\\
&\times\bigg(\prod_{k=3}^n\int\frac{d^3q^{\prime }_k}{q^{\prime0}_k}G_{\Delta_k}(\hat{q}_k;q^{\prime}_k)\bigg)(2\pi)^4\delta^{(4)}(p-\sum_{k=3}^nq^{\prime}_k)\mathcal{T}_{2\rightarrow n-2}(q^{\prime}_i)\;,
\end{split}
\end{align}
where $Y^{\prime\mu}\equiv2(-\hat{q}_{2}\cdot\hat{p})\hat{p}^{\mu}-\hat{q}_{2}^{\mu}$\;.
For $n\geq3$-point shadow celestial amplitudes, the OPE limit can be studied by the generalized optical theorems. We assume $\mathcal{M}_{2\rightarrow n-2}(q^{\prime}_1,q^{\prime}_2,\cdots,q^{\prime}_n)$ is at tree-level. Using the generalized optical theorem \eqref{eq:optical} and noting that there are only single-particle intermediate states that can appear in tree diagrams, we get \footnote{Here and throughout this paper, we assume that conformal dimensions $\Delta_i$ are real when taking the imaginary part of $\widetilde{\mathcal{A}}^{\Delta_i}_{2\rightarrow n-2}(z_i)$.}
\begin{align}
\begin{split}
\lim_{\hat{q}_1\rightarrow\hat{q}_2}\text{Im}\,\widetilde{\mathcal{A}}^{\Delta_i}_{2\rightarrow n-2}&=\int d^{4}p\int \frac{D^2\hat{q}^{\prime}_2\;2^{\Delta_1+\Delta_{2}-1}|p|^{2-2\Delta_1-2\Delta_2}}{(-\hat{q}_2\cdot\hat{q}^{\prime}_2)^{\Delta_2}(-\hat{q}^{\prime}_2\cdot p)^{2-\Delta_1-\Delta_2}(-\hat{q}^{\prime}_2\cdot Y^{\prime})^{\Delta_1}}\bigg(\prod_{k=3}^n\int\frac{d^3q^{\prime }_k}{q^{\prime0}_k}G_{\Delta_k}(\hat{q}_k;q^{\prime}_k)\bigg)\\
&\times \sum_{X}\int\frac{d^3p_X}{(2\pi)^3}\frac{1}{2E_X}(2\pi)^4\delta^{(4)}(p-p_X)\mathcal{T}_{2\rightarrow X}(p^2)\mathcal{M}^{*}_{X\rightarrow n-2}(p_X,q^{\prime}_i)\;.
\end{split}
\end{align}
Here $X$ labels all possible physical single-particle states that satisfy the on-shell condition $p_X^2=-m_X^2$. With the help of the delta function, we evaluate the integral over $p$, leading to
\begin{align}
\begin{split}
\lim_{\hat{q}_1\rightarrow\hat{q}_2}\text{Im}\,\widetilde{\mathcal{A}}^{\Delta_i}_{2\rightarrow n-2}=&2^{\Delta_{1}+\Delta_{2}}\pi\sum_X\int\frac{d^3\hat{p}_X}{2\hat{p}^0_X}\int D^2\hat{q}^\prime_2\frac{m_X^{2-\Delta_1-\Delta_2}\mathcal{T}_{2\rightarrow X}(m_X^2)}{(-\hat{q}_2\cdot\hat{q}^{\prime}_2)^{\Delta_2}(-\hat{q}^{\prime}_2\cdot\hat{p}_X)^{2-\Delta_1-\Delta_2}(-\hat{q}^{\prime}_2\cdot Y^{\prime})^{\Delta_{1}}}\\
&\qquad\times\bigg(\prod_{k=3}^n\int\frac{d^3q^{\prime }_k}{q^{\prime0}_k}G_{\Delta_k}(\hat{q}_k;q^{\prime}_k)\bigg)\mathcal{M}^{*}_{X\rightarrow n-2}(m_X\hat{p}_X,q^{\prime}_i)\;.
\end{split}
\end{align}
After evaluating the integral over $\hat{q}^{\prime}_2$ by using \eqref{eq:ConformalI1}, we get
\begin{align}\label{eq:OPEO}
\begin{split}
\lim_{\hat{q}_1\rightarrow\hat{q}_2}\text{Im}\,\widetilde{\mathcal{A}}^{\Delta_i}_{2\rightarrow n-2}=&\sum_X\frac{\pi^2\Gamma[1-\Delta_1]\Gamma[1-\Delta_2]}{\Gamma[2-\Delta_1-\Delta_2]}m_X^{2-\Delta_1-\Delta_2}\mathcal{T}_{2\rightarrow X}(m_X^2)\widetilde{\mathcal{A}}^{*\Delta_1+\Delta_2,\Delta_i}_{X\rightarrow n-2}\\
=&\frac{1}{2}\sum_X\mathcal{C}_{\Delta_1,\Delta_2,\Delta_1+\Delta_2}(m_X)D^{\Delta_1+\Delta_2,\Delta_1+\Delta_2}(m_X)\widetilde{\mathcal{A}}^{*\Delta_1+\Delta_2,\Delta_i}_{X\rightarrow n-2}\;,
\end{split}
\end{align}
where $D^{\Delta\Delta}(m)=\frac{(\Delta-1)m^2}{(2\pi)^5}$ is the inverse of coefficient of two-point massive celestial amplitude \cite{Atanasov:2021cje} and $\mathcal{C}_{\Delta_1,\Delta_2,\Delta_1+\Delta_2}(m_X)$ is given in \eqref{eq:C3pt}

Some comments are made here. The leading OPE behaviour \eqref{eq:OPEO} works only when the $(n-1)$-point shadow celestial amplitudes $\widetilde{\mathcal{A}}^{\Delta_1+\Delta_2,\Delta_i}_{X\rightarrow n-2}(m_X;z_i)$ are well-defined. This demands that the integrals appearing in the computation of the $(n-1)$-point shadow celestial amplitudes $\widetilde{\mathcal{A}}^{\Delta_1+\Delta_2,\Delta_i}_{X\rightarrow n-2}(m_X;z_i)$ must converge. For example, in the case of $n=4$, the leading OPE behaviour \eqref{eq:OPEO} works only when $\text{Re}(\Delta_3+\Delta_4-\Delta_1-\Delta_2)>0$, beyond which the integrals appearing in $\widetilde{\mathcal{A}}^{\Delta_1+\Delta_2,\Delta_3,\Delta_4}_{1\rightarrow 2}(m_X;z_i)$ are divergent. Although one can still define $\widetilde{\mathcal{A}}^{\Delta_1+\Delta_2,\Delta_3,\Delta_4}_{1\rightarrow 2}(m_X;z_i)$ by analytical continuation when $\text{Re}(\Delta_3+\Delta_4-\Delta_1-\Delta_2)\leq0$, \eqref{eq:OPEO} is no longer dominant in this region.

\subsection{OPE analysis for a special class of Feynman diagrams}\label{ssec:OPES}

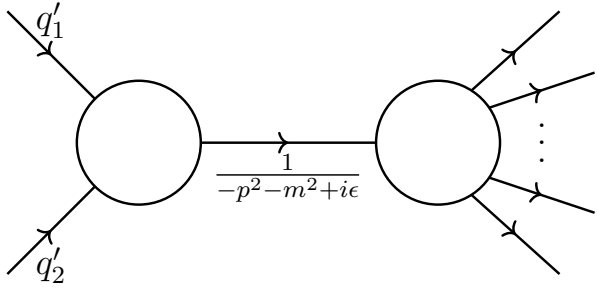
\begin{figure}[!t]
\centering
\resizebox{8cm}{!}{%
\begin{tikzpicture}[thick]
\begin{scope}
\draw[-,postaction={decorate,decoration={markings,mark=at position 0.5 with {\arrow{>}}}}] (-1.5,1.5)--(-0.5,0.5) node at (-1,1.4){$q^{\prime}_1$};
\draw[-,postaction={decorate,decoration={markings,mark=at position 0.5 with {\arrow{>}}}}] (-1.5,-1.5)--(-0.5,-0.5) node at (-1,-1.4){$q^{\prime}_2$};
\draw[-] (0,0) circle (0.707);
\draw[-,postaction={decorate,decoration={markings,mark=at position 0.5 with {\arrow{>}}}}] (0.707,0)--(2.707,0) node at (1.7,-0.4){$\frac{1}{-p^2-m^2+i\epsilon}$};
\draw[-] (3.414,0) circle (0.707);
\draw[-,postaction={decorate,decoration={markings,mark=at position 0.5 with {\arrow{>}}}}] (3.8,0.6)--(4.8,1.5);
\draw[-,postaction={decorate,decoration={markings,mark=at position 0.5 with {\arrow{>}}}}] (4.0,0.4)--(5.2,0.8);
\node at (4.6,0.2){$.$};
\node at (4.6,0.0){$.$};
\node at (4.6,-0.2){$.$};
\draw[-,postaction={decorate,decoration={markings,mark=at position 0.5 with {\arrow{>}}}}] (3.8,-0.6)--(4.8,-1.5);
\draw[-,postaction={decorate,decoration={markings,mark=at position 0.5 with {\arrow{>}}}}] (4.0,-0.4)--(5.2,-0.8);
\end{scope}
\end{tikzpicture}
}
\caption{A particular class of Feynman diagrams which contain a massive propagator $\frac{1}{-p^2-m^2+i\epsilon}$ with $p=q^{\prime}_1+q^{\prime}_2$.}
\label{fig:Feynman}
\end{figure}

In the previous subsection, we used the generalized optical theorem to decompose a $n$-point scattering amplitude into a 3-point and a $(n-1)$-point scattering amplitudes.
The shortcoming is that we only obtained the OPE limit of the imaginary part of the shadow celestial amplitude $\widetilde{\mathcal{A}}^{\Delta_i}_{2\rightarrow n-2}$.
In this subsection, we will focus on a special class of Feynman diagrams shown in Figure \ref{fig:Feynman}. It allows us to derive a formula for the OPE limit of the shadow celestial amplitude without taking the imaginary part. 
We note that the scattering amplitudes of the class of Feynman diagrams in Figure \ref{fig:Feynman} take the form
\begin{align}
\mathcal{M}_{2\rightarrow n-2}(q^{\prime}_i)=(2\pi)^4\delta^{(4)}(p-\sum_{k=3}^nq^{\prime}_k)i\mathcal{T}_{2\rightarrow 1}(p^2)\frac{1}{-p^2-m^2}i\mathcal{T}_{1\rightarrow n-2}(p,q^{\prime}_i)\;,
\end{align}
where we defined $p=q^{\prime}_1+q^{\prime}_2$. Plugging the above express into \eqref{eq:OPE1} and changing integral variables from $p^{\mu}$ to $M\hat{p}^{\mu}$ with $\hat{p}\cdot\hat{p}=-1$ then lead to
\begin{align}
\begin{split}
\lim_{\hat{q}_1\rightarrow\hat{q}_2}\widetilde{\mathcal{A}}^{\Delta_i}_{2\rightarrow n-2}&=-2^{\Delta_1+\Delta_{2}}\int_0^{+\infty}dM\int\frac{d^{3}\hat{p}}{\hat{p}^0}\int\frac{D^2\hat{q}^{\prime}_2M^{3-\Delta_1-\Delta_2}}{(-\hat{q}_2\cdot\hat{q}^{\prime}_2)^{\Delta_2}(-\hat{q}^{\prime}_2\cdot\hat{p})^{2-\Delta_1-\Delta_2}(-\hat{q}^{\prime}_{2}\cdot Y^{\prime})^{\Delta_1}}\\
&\times\bigg(\prod_{k=3}^n\int\frac{d^3q^{\prime}_k}{q^{\prime0}_k}G_{\Delta_k}(\hat{q}_k;q^{\prime}_k)\bigg)\mathcal{T}_{2\rightarrow 1}(M^2)\frac{1}{M^2-m^2}\mathcal{M}_{1\rightarrow n-2}(p,q^{\prime}_i)\;.
\end{split}
\end{align}
Performing the integral over $\hat{q}^{\prime}_2$, we obtain
\begin{align}\label{eq:OPES}
\begin{split}
\lim_{\hat{q}_1\rightarrow\hat{q}_2}\widetilde{\mathcal{A}}^{\Delta_i}_{2\rightarrow n-2}=&\frac{-2\pi\Gamma[1-\Delta_1]\Gamma[1-\Delta_2]}{\Gamma[2-\Delta_1-\Delta_2]}\int_0^{+\infty}dM\frac{M^{3-\Delta_1-\Delta_2}\mathcal{T}_{2\rightarrow 1}(M^2)}{M^2-m^2}\widetilde{\mathcal{A}}_M^{\Delta_1+\Delta_2,\Delta_i}\\
=&-\frac{1}{\pi}\int_0^{+\infty}dM\frac{M\mathcal{C}_{\Delta_1,\Delta_2,\Delta_1+\Delta_2}(M)D^{\Delta_1+\Delta_2,\Delta_1+\Delta_2}(M)}{M^2-m^2}\widetilde{\mathcal{A}}_M^{\Delta_1+\Delta_2,\Delta_i}\;.
\end{split}
\end{align}
Thus, for the particular class of diagrams in Figure \ref{fig:Feynman}, the leading OPE limit of a $n$-point massless shadow celestial amplitudes are given by a superposition of $(n-1)$-point shadow celestial amplitudes over a range of mass $M\in[0,\infty)$. \eqref{eq:OPES} leads to the leading order operator product expansion
\begin{align}\label{eq:OPES1}
\mathcal{O}_1(z_1)\mathcal{O}_2(z_2)\sim-\frac{1}{\pi}\int_0^{+\infty}dM\frac{M\mathcal{C}_{\Delta_1,\Delta_2,\Delta_1+\Delta_2}(M)D^{\Delta_1+\Delta_2,\Delta_1+\Delta_2}(M)}{M^2-m^2}\Phi^{\Delta_1+\Delta_2}_M(z_2)\;.
\end{align}

We stress here again that \eqref{eq:OPES} works only when the $(n-1)$-point shadow celestial amplitudes $\widetilde{\mathcal{A}}^{\Delta_1+\Delta_2,\Delta_i}_{M}$ are well-defined.

\subsection{OPE analysis for four-point shadow celestial amplitudes}

As an example, we consider the OPE behaviour of four-point shadow celestial amplitudes $\widetilde{\mathcal{A}}^{\Delta_i}_{2\rightarrow2}$ involving two incoming and two outgoing massless scalars with a massive scalar exchange. Since only $s$-channel amplitude has an imaginary part and is in our particular class of diagrams shown in Figure \ref{fig:Feynman}, we only focus on the $s$-channel amplitude and take the OPE limit $\hat{q}_1\rightarrow\hat{q}_2$. The OPE limit $\hat{q}_3\rightarrow\hat{q}_4$ can be derived in a similar way. In this specific case, \eqref{eq:OPEO} takes the form as 
\begin{align}\label{eq:OPEO4pt}
\begin{split}
\lim_{\hat{q}_1\rightarrow\hat{q}_2}\text{Im}\,\widetilde{\mathcal{A}}_{s}^{\Delta_i}(z_i)=&\frac{1}{2}\mathcal{C}_{\Delta_1,\Delta_2,\Delta_1+\Delta_2}(m)D^{\Delta_1+\Delta_2,\Delta_1+\Delta_2}(m)\widetilde{\mathcal{A}}^{\Delta_3,\Delta_4,\Delta_1+\Delta_2}_m\;.
\end{split}
\end{align}
Here $m$ denotes the mass of the exchange operator. On the other hand, in this case, \eqref{eq:OPES} becomes
\begin{align}
\begin{split}
\lim_{\hat{q}_1\rightarrow\hat{q}_2}\widetilde{\mathcal{A}}^{\Delta_i}_{2\rightarrow 2}
=&-\frac{1}{\pi}\int_0^{+\infty}dM\frac{M\mathcal{C}_{\Delta_1,\Delta_2,\Delta_1+\Delta_2}(M)D^{\Delta_1+\Delta_2,\Delta_1+\Delta_2}(M)}{M^2-m^2}\widetilde{\mathcal{A}}_M^{\Delta_1+\Delta_2,\Delta_3,\Delta_4}\;.
\end{split}
\end{align}
We note that in the present case $\widetilde{\mathcal{A}}_M^{\Delta_1+\Delta_2,\Delta_i}(z_i)$ takes the form as
\begin{align}
\begin{split}
\widetilde{\mathcal{A}}^{\Delta_1+\Delta_2,\Delta_3,\Delta_4}_{M}=&-\frac{g}{2^{\Delta_3+\Delta_4-4}}\int\frac{d^2z^{\prime}_3}{|z_3-z^{\prime}_3|^{2\Delta_3}}\frac{d^2z^{\prime}_4}{|z_4-z^{\prime}_4|^{2\Delta_4}}\int\frac{d^{3}\hat{p}}{\hat{p}^0}\frac{1}{(-\hat{q}_{2}\cdot\hat{p})^{\Delta_{1}+\Delta_2}}\\
&\qquad\times\int_{0}^{\infty}d\omega_3\,\omega_3^{1-\Delta_3}\int_{0}^{\infty}d\omega_4\,\omega_4^{1-\Delta_4}(2\pi)^4\delta^{(4)}(M\hat{p}-\omega_3\hat{q}^{\prime}_3-\omega_4\hat{q}^{\prime}_4)\;.
\end{split}
\end{align}
Rescaling $\omega_3\rightarrow M\omega_3$ and $\omega_4\rightarrow M\omega_4$ leads to
\begin{align}
\begin{split}
\widetilde{\mathcal{A}}^{\Delta_1+\Delta_2,\Delta_3,\Delta_4}_{M}(z_i)=&\frac{-gM^{-\Delta_3-\Delta_4-2}}{2^{\Delta_3+\Delta_4-4}}\int\frac{d^2z^{\prime}_3}{|z_3-z^{\prime}_3|^{2\Delta_3}}\frac{d^2z^{\prime}_4}{|z_4-z^{\prime}_4|^{2\Delta_4}}\int\frac{d^{3}\hat{p}}{\hat{p}^0}\frac{1}{(-\hat{q}_{2}\cdot\hat{p})^{\Delta_{1}+\Delta_2}}\\
&\qquad\times\int_{0}^{\infty}d\omega_3\omega_3^{1-\Delta_3}\int_{0}^{\infty}d\omega_4\omega_4^{1-\Delta_4}(2\pi)^4\delta^{(4)}(\hat{p}-\omega_3\hat{q}^{\prime}_3-\omega_4\hat{q}^{\prime}_4)\\
=&M^{-\Delta_3-\Delta_4}\widetilde{\mathcal{A}}^{\Delta_1+\Delta_2,\Delta_3,\Delta_4}_{M=1}(z_i)\;.
\end{split}
\end{align}
Plugging the above expression into \eqref{eq:OPES} then leads to
\begin{align}\label{eq:OPES4pt}
\begin{split}
\lim_{\hat{q}_1\rightarrow\hat{q}_2}\widetilde{\mathcal{A}}^{\Delta_i}_{s}=&-\frac{1}{\pi}\int_0^{+\infty}dM\frac{M^{-\beta-1}\mathcal{C}_{\Delta_1,\Delta_2,\Delta_1+\Delta_2}(1)D^{\Delta_1+\Delta_2,\Delta_1+\Delta_2}(1)}{M^2-m^2}\widetilde{\mathcal{A}}^{\Delta_1+\Delta_2,\Delta_3,\Delta_4}_{M=1}(z_i)\\
=&\frac{e^{-i\pi\beta/2}}{2\sin(\frac{-\pi\beta}{2})}\mathcal{C}_{\Delta_1,\Delta_2,\Delta_1+\Delta_2}(m)D^{\Delta_1+\Delta_2,\Delta_1+\Delta_2}(m)\widetilde{\mathcal{A}}^{\Delta_1+\Delta_2,\Delta_3,\Delta_4}_{m}\;,
\end{split}
\end{align}
where we defined $\beta=4-\Delta_1-\Delta_2-\Delta_3-\Delta_4$. We stress here that \eqref{eq:OPEO4pt} and \eqref{eq:OPES4pt} hold only when $\widetilde{\mathcal{A}}^{\Delta_1+\Delta_2,\Delta_3,\Delta_4}_{m}$ is finite, which constrains the conformal dimensions by $\text{Re}(\Delta_1+\Delta_2-\Delta_3-\Delta_4)<0$. 

By further expanding \eqref{eq:OPES4pt} in power series of $\hat{q}_{34}$ and using \eqref{eq:OPE3pt}, we find
\begin{align}\label{eq:OPE12then34}
\begin{split}
\lim_{\hat{q}_1\rightarrow\hat{q}_2}\widetilde{\mathcal{A}}^{\Delta_i}_{s}=&\frac{e^{-i\pi\beta/2}}{2\sin(\frac{-\pi\beta}{2})}\mathcal{C}_{\Delta_1,\Delta_2,\Delta_1+\Delta_2}(m)D^{\Delta_1+\Delta_2,\Delta_1+\Delta_2}(m){\mathcal{C}}_{\Delta_3,\Delta_4,\Delta_1+\Delta_2}(m)
\\
&\times\hat{q}_{24}^{-\Delta_1-\Delta_2}\hat{q}_{34}^{\frac{\Delta_1+\Delta_2-\Delta_3-\Delta_4}{2}}(1+{\mathcal{O}}(\hat{q}_{34}))
\end{split}
\end{align}
In the next section, we will compute the $s$-channel four-point shadow celestial amplitude and derive its conformal block expansion. We will find agreement between the conformal block expansion and the OPE expansion \eqref{eq:OPE12then34} when $\text{Re}(\Delta_1+\Delta_2-\Delta_3-\Delta_4)<0$.

\section{Examples of shadow celestial amplitudes} \label{sec:5}

\subsection{Three-point shadow celestial amplitudes}
 
 We first consider the shadow celestial amplitudes for two incoming massless scalars and one outgoing massive scalar. The celestial amplitudes $\mathcal{A}_{2\rightarrow 1}^{\Delta_i}$ at tree level is \cite{Lam:2017ofc}
\begin{align}\label{eq:OCA3pt}
\mathcal{A}_{2\rightarrow 1}^{\Delta_i}(z_i)=-g\frac{\pi^4m^{\Delta_1+\Delta_2-4}}{2^{\Delta_1+\Delta_2-5}}\frac{B(\frac{\Delta_1+\Delta_{12}}{2},\frac{\Delta_1-\Delta_{12}}{2})}{|z_{12}|^{\Delta_1+\Delta_2-\Delta_3}|z_{13}|^{\Delta_1-\Delta_2+\Delta_3}|z_{23}|^{\Delta_{2}+\Delta_3-\Delta_1}}\;,
\end{align}
where $B(a,b)\equiv\Gamma[a]\Gamma[b]/\Gamma[a+b]$ is the beta-function. With the help of \eqref{eq:identity1}, we compute the three-point shadow celestial amplitudes $\widetilde{\mathcal{A}}^{\Delta_i}_{2\rightarrow 1}$. We get
\begin{align}\label{eq:3pt}
\widetilde{\mathcal{A}}_{2\rightarrow 1}^{\Delta_i}(z_i)=\frac{C_{\Delta_1,\Delta_2,\Delta_3}}{|z_{12}|^{\Delta_1+\Delta_2-\Delta_3}|z_{13}|^{\Delta_1-\Delta_2+\Delta_3}|z_{23}|^{\Delta_{2}+\Delta_3-\Delta_1}}\;,
\end{align}
where the coefficient $C_{\Delta_1,\Delta_2,\Delta_3}$ is given in \eqref{eq:C3pt1} with $\mathcal{T}_{2\rightarrow 1}(m^2)=-g$ and $z_{ij}\equiv z_i-z_j$ and $\bar{z}_{ij}\equiv\bar{z}_i-\bar{z}_j$. We mention here that the integral in the shadow transformation from \eqref{eq:OCA3pt} to \eqref{eq:3pt} converges only when $\text{Re}(\Delta_1+\Delta_2-\Delta_3)>0$. However, with \eqref{eq:3pt} in hand, we can analytically continue the conformal dimensions to the whole complex plane excluding the poles of $C_{\Delta_1,\Delta_2,\Delta_3}$. 

The constrains imposed by translation symmetries \eqref{eq:TS} on the shadow celestial amplitude $\widetilde{\mathcal{A}}^{\Delta_i}_{2\rightarrow 1}$ are
\begin{align}\label{eq:TS3pt}
\begin{split}
%&\bigg[(\Delta_1+\Delta_{23}-1)^2(\Delta_3-1)e^{-\partial_{\Delta_{2}}}-2(\Delta_2-1)^2(\Delta_{12}+\Delta_3-1)me^{-\partial_{\Delta_{3}}}\bigg]C_{\Delta_1,\Delta_2,\Delta_3}(m)=0\;,\\
&\bigg[(\Delta_{12}+\Delta_{3}-1)(\sum_{i=1}^3\Delta_i-3)^2e^{-\partial_{\Delta_{1}}}-8(\Delta_{1}-1)^{2}\Delta_{3}me^{\partial_{\Delta_{3}}}\bigg]C_{\Delta_1,\Delta_2,\Delta_3}(m)=0\;,\\
&\bigg[(-\Delta_{12}+\Delta_{3}-1)(\sum_{i=1}^3\Delta_{i}-3)^2e^{-\partial_{\Delta_{2}}}-8(\Delta_{2}-1)^{2}\Delta_{3}me^{\partial_{\Delta_{3}}}\bigg]C_{\Delta_1,\Delta_2,\Delta_3}(m)=0\;,\\
&\bigg[(\Delta_{12}-\Delta_{3}+1)(\Delta_{12}+\Delta_3-1)(\sum_{i=1}^3\Delta_i -3)^2 e^{-\partial_{\Delta_{3}}} \\ 
& \qquad \qquad \qquad \qquad +4\Delta_3(\Delta_1+\Delta_2-\Delta_3-1)^2(\Delta_3-1) e^{\partial_{\Delta_{3}}}\bigg]C_{\Delta_1,\Delta_2,\Delta_3}(m)=0\;.
\end{split}        
\end{align}
It is then straightforward to confirm that the coefficient \eqref{eq:C3pt1} satisfies the above equality. One can also check that the OPE behaviour of \eqref{eq:3pt} agrees with our analysis in \eqref{eq:OPE3pt} when $\Delta_3\neq \Delta_1+\Delta_2+2n$ and \eqref{eq:OPE3pt1} when $\Delta_3=\Delta_1+\Delta_2+2n$ after setting $\mathcal{T}_{2\rightarrow 1}(m^2)=-g$.

\subsection{Four-point shadow celestial amplitudes}

In this subsection, we will compute the four-point shadow celestial amplitudes $\widetilde{\mathcal{A}}^{\Delta_i}_{12\rightarrow34}$ of four external massless scalars and one exchange massive scalar, and derive its conformal block expansion in the $12-34$ channel. To achieve this, we start with the corresponding  celestial amplitudes $\mathcal{A}^{\Delta_i}_{12\rightarrow34}$ with the explicit formulae given in \eqref{eq:A}, \eqref{eq:I1234} and \eqref{eq:4pt}, and perform the shadow transformation for each external operator.

%%%%%%%%%%%%%%%%%%%%%%%%%

\subsubsection{Shadow transformation for $\mathcal{O}_1$}

To get the shadow celestial amplitudes $\widetilde{\mathcal{A}}^{\Delta_i}_{s}$, we first perform the shadow transformation on the operator $\mathcal{O}_1$, and denote the result by $\widehat{{\cal A}}^{\Delta_i}_s$.\footnote{See \cite{Fan:2021isc,Fan:2021pbp} for performing the shadow transformation on one of the operators in four-point gluon amplitudes.} Using the conformal symmetry, we can fix three of the four coordinates, leading to 
\begin{align}\label{eq:GaugeFixing}
z_1=\bar{z}_1=\chi\in(1,+\infty)\;,\hspace{0.5cm}z_2=\bar{z}_2=0\;,\hspace{0.5cm}z_3=\bar{z}_3=+\infty\;,\hspace{0.5cm}z_4=\bar{z}_4=1\;.
\end{align}
In this coordinate configuration, $I_{12-34}$ in \eqref{eq:I1234} takes the form as
\begin{align}
    I_{12-34}=|\chi|^{-2h_{1}-2h_{2}}|\chi-1|^{2h_{34}-2h_{12}}\dfrac{\left|z_{24}\right|^{2h_{12}}|z_{3}|^{-2h_{34}}}{|z_{34}|^{2h_{3}+2h_{4}}}\;,
\end{align}
where $\Delta_i=2h_i$.
Together with \eqref{eq:A} and \eqref{eq:4pt}, we simplify the shadow integral as
\begin{align}\label{eq:FirstShadow}
\begin{split}
\widehat{{\cal A}}_{s}^{\widetilde{\Delta}_{1},\Delta_{2},\Delta_{3},\Delta_{4}}=&2^{\Delta_{1}}\int_{1}^{+\infty}\frac{d^{2}z_{1}}{(z_{1}-z_{1}^{\prime})^{2\widetilde{h}_{1}}(\bar{z}_{1}-\bar{z}_{1}^{\prime})^{2\widetilde{h}_{1}}}I_{12-34}f_{s}(\chi)\\
=&2^{\Delta_{1}}N_{gm}(\beta)\dfrac{\left|z_{24}\right|^{2h_{12}}|z_{3}|^{-2h_{34}}}{|z_{34}|^{2h_{3}+2h_{4}}}\int_{1}^{\infty}\frac{d^{2}\chi\delta(\chi-\bar{\chi})\chi^{2\widetilde{h}_{1}-2h_{2}}(\chi-1)^{h_{34}-h_{12}}}{(\chi-\chi^{\prime})^{2\widetilde{h}_{1}}(\bar{\chi}-\bar{\chi}^{\prime})^{2\widetilde{h}_{1}}}\\
=&2^{\Delta_{1}}N_{gm}(\beta)\dfrac{\left|z_{24}\right|^{2h_{12}}|z_{3}|^{-2h_{34}}}{|z_{34}|^{2h_{3}+2h_{4}}}\int_{0}^{1}\frac{d\chi\chi^{h_{1^{\prime}}+h_{2}-h_{34}-1}(1-\chi)^{h_{34}-h_{12}}}{(1-\chi\chi^{\prime})^{2h_{1^{\prime}}}(1-\chi\bar{\chi}^{\prime})^{2\bar{h}_{1^{\prime}}}}\;,
\end{split}
\end{align}
where $\widetilde \Delta_1=2-\Delta_1$ and $\widetilde{h}_{1}=1-h_1$, and we defined the new conformal cross-ratios $\chi^{\prime}$ and $\bar{\chi}^{\prime}$ as
\begin{align}
\chi^{\prime}=\dfrac{z_{1^{\prime}2}z_{34}}{z_{1^{\prime}3}z_{24}}\;,\hspace{1cm}\bar{\chi}^{\prime}=\dfrac{\bar{z}_{1^{\prime}2}\bar{z}_{34}}{\bar{z}_{1^{\prime}3}\bar{z}_{24}}\;,\hspace{0.5cm}z_{i^{\prime}j}\equiv z^{\prime}_i-z_j\;.
\end{align}
The above integral can be evaluated by using the integral representation of the Appell function $F_1$:
\begin{align}\label{eq:Appell}
F_1\left(\left.\begin{array}{c}a; b_1, b_2 \\ c \end{array}\right.\bigg|x,y\right)=\frac{\Gamma[c]}{\Gamma[a]\Gamma[c-a]}\int_{0}^1dt t^{a-1}(1-t)^{c-a-1}(1-tx)^{-b_1}(1-ty)^{-b_2}\;,
\end{align}
leading to
\begin{align}
\begin{split}
\widehat{{\cal A}}^{\widetilde{\Delta}_1,\Delta_2,\Delta_3,\Delta_4}_s=&2^{\Delta_1}N_{gm}(\beta)\dfrac{\left|z_{24} \right|^{2h_{12}}|z_3|^{-2h_{34}}}{|z_{34}|^{2h_3+2h_4}}B\left(\widetilde{h}_{1}+h_2-h_{34},\widetilde{h}_{1}+h_2+h_{34}\right)\\
&\qquad\times  F_1\left(\left.\begin{array}{c}\widetilde h_1+h_2-h_{34}; 2\widetilde h_1, 2\widetilde h_1 \\ 2\widetilde h_1+2h_2 \end{array}\right.\bigg|\chi^{\prime},\bar{\chi}^{\prime}\right)\;.
\end{split}
\end{align}
Using the conformal symmetry, one can unfix the coordinates \eqref{eq:GaugeFixing}, and reach the following expression
\begin{align}
\begin{split}
\widehat{{\cal A}}^{\Delta_i}_{s}(z_i)=&I_{12-34}\mathcal{G}_{s}(\chi)\;,
\end{split}
\end{align}
where we rename all of the primed (tilde) variables by the corresponding unprimed (untilde) variables to simplify the notation and $\mathcal{G}_{s}(\chi)$ is
\begin{align}
\begin{split}
\mathcal{G}_{s}(\chi)=&2^{2-\Delta_1}N_{gm}(\beta)B\left(h_{1}+h_2-h_{34},h_{1}+h_2+h_{34}\right)\\
&\times (\chi\bar{\chi})^{h_{1}+h_2}|1-\chi|^{2(h_{13}-h_{24})}F_1\left(\left.\begin{array}{c}h_{1}+h_2-h_{34}; 2h_{1}, 2h_{1}\\ 2h_{1}+2h_2 \end{array}\right.\bigg|\chi,\bar{\chi}\right)\;,
\end{split}
\end{align}
where we have renamed $\widetilde\Delta_1$, $\widetilde{h}_1$ by $\Delta_1$, $h_1$, and in this convention $\beta=-\Delta_1+\Delta_2+\Delta_3+\Delta_4-2$. To find the conformal block expansion in the $12-34$ channel, we use the Burchnall-Chaundy expansion for the Appell function
\begin{align}\label{eq:AppellIdentity}
\begin{split}
F_1\left(\left.\begin{array}{c}a; b_1, b_2 \\ c \end{array}\right.\bigg|x,y\right)=&\sum_{n=0}^{+\infty}\frac{(a)_n(b_1)_n(b_2)_n(c-a)_n}{n!(c+n-1)_n(c)_{2n}}x^ny^n\\
&\qquad\times{}_2F_1\left(\left.\begin{array}{c}a+n, b_1+n \\ c+2n \end{array}\right.\bigg|x\right){}_2F_1\left(\left.\begin{array}{c}a+n, b_2+n\\ c+2n \end{array}\right.\bigg|y\right)\;,
\end{split}
\end{align}
leading to
\begin{align}
\begin{split}
\mathcal{G}_{s}(\chi)=&2^{2-\Delta_1}N_{gm}(\beta)\sum_{n=0}^{+\infty}\frac{B(h-h_{34},h+h_{34})(h+h_{12}-n)_n(h+h_{12}-n)_n}{n!(2h-n-1)_n}\\
&\times|1-\chi|^{2(h_{13}-h_{24})}(\chi\bar{\chi})^{h}{}_2F_1\left(\left.\begin{array}{c}h-h_{34}, h+h_{12}\\ 2h\end{array}\right.\bigg|\chi\right){}_2F_1\left(\left.\begin{array}{c}\bar{h}-\bar{h}_{34}, \bar{h}+\bar{h}_{12}\\ 2\bar{h}\end{array}\right.\bigg|\bar{\chi}\right)\;,
\end{split}
\end{align}
where we defined 
\begin{align}
h=\bar{h}=h_{1}+h_2+n\;,\hspace{1cm}n\in\mathbb{Z}_{\geq0}\;.
\end{align}
Finally, after using the following identity
\begin{align}\label{eq:2F101}
{}_2F_1\left(\left.\begin{array}{c}a, b \\ c \end{array}\right.\bigg|x\right)=(1-x)^{c-a-b}{}_2F_1\left(\left.\begin{array}{c}c-a, c-b \\ c \end{array}\right.\bigg|x\right)\;,
\end{align}
we get the conformal block expansion of $\mathcal{G}_{s}(\chi)$ in $12-34$ channel,
\begin{align}\label{eq:BE1st}
\begin{split}
\mathcal{G}_{s}(\chi)=&N_{gm}(\beta)\sum_{n=0}^{+\infty}\frac{B(h_{1+2}-h_{34}+n,h_{1+2}+h_{34}+n)\Gamma[2h_1+n]^2}{2^{\Delta_1-2}n!\Gamma[2h_1]\Gamma[1-2h_1](2h_{1+2}+n-1)_n}G^{h_i,\bar{h}_i}_{h_{1+2}+n,h_{1+2}+n}(\chi)\;.
\end{split}
\end{align}
Here, we introduced the notation $h_{i+j}\equiv h_i+h_j$ and the conformal block in $s$-channel is given by
\begin{align}
G^{h_i,\bar{h}_i}_{h,\bar{h}}(\chi)=\chi^h{}_2F_1\left(\left.\begin{array}{c}h-h_{12}, h+h_{34} \\ 2h \end{array}\right.\bigg|\chi\right)\bar{\chi}^{\bar{h}}{}_2F_1\left(\left.\begin{array}{c}\bar{h}-\bar{h}_{12}, \bar{h}+\bar{h}_{34} \\ 2\bar{h} \end{array}\right.\bigg|\bar{\chi}\right)\;.
\end{align}

\subsubsection{Shadow transformation for $\mathcal{O}_2$, $\mathcal{O}_3$ and $\mathcal{O}_4$}

To get the shadow celestial amplitude $\widetilde{\mathcal{A}}^{\Delta_i}_{s}$, we still need perform the shadow transformations on the remaining three operators in $\widehat{\mathcal{A}}^{\Delta_i}_{s}$.  However, computing the shadow integral as we did for $\mathcal{O}_1$ now becomes extremely hard. Instead of computing the shadow integral directly, in this subsection, we will use the conformal partial wave reviewed in the Appendix \ref{sec:PW} to get the conformal block expansion of $\widetilde{\mathcal{A}}^{\Delta_i}_{s}$. Specifically, we first derive the conformal partial expansion of $\widehat{\mathcal{A}}^{\Delta_i}_{s}$ from the block expansion \eqref{eq:BE1st}. Then we compute the conformal partial wave expansion of $\widetilde{\mathcal{A}}^{\Delta_i}_{s}$ by virtue of \eqref{eq:It=I}. Finally, we derive the desired conformal block expansion by closing the contour to the right-half plane. 

From the block expansion \eqref{eq:BE1st} and the symmetry property \eqref{eq:I=I}, we can derive the spectral density $\widehat{\rho}^{\Delta_i}_{h,\bar{h}}$ for $\widehat{\mathcal{A}}^{\Delta_i}_{s}$,
%
%\begin{align}
%\frac{\widehat{\rho}_{h,\bar{h}}^{\Delta_i}}{n_{h,\bar{h}}}=&N_{gm}(\beta)\frac{-\Gamma[1-2h_1]\Gamma[h+h_{12}]\Gamma[h_{1+2}-h]\Gamma[h+h_{1+2}-1]\Gamma[1-h+h_{34}]\Gamma%[1-h-h_{34}]}{2^{\Delta_1-2}\Gamma[2h_1]\Gamma[2h-1]\Gamma[1-2h]\Gamma[1-h_{12}-h]}\;,
%\end{align}
%
%or
%
\begin{align}\label{eq:FirstRho}
\begin{split}
K^{\Delta_3,\Delta_4}_{1-h,1-\bar{h}}\frac{\widehat{\rho}_{h,\bar{h}}^{\widetilde{\Delta}_1,\Delta_2,\Delta_3,\Delta_4}}{n_{h,\bar{h}}}=& -\frac{g^2\pi^9m^{\beta-4}}{2^{\beta-6}}\frac{e^{i\pi\beta/2}}{\sin \pi\beta/2}\frac{a^{\Delta_1,\Delta_2}_{\Delta}}{2^{-\Delta_1}}\\
&\times B(h+h_{12},h-h_{12})B(h-h_{34},h+h_{34})D^{\Delta\Delta}(m)\;,
\end{split}
\end{align}
where $\Delta=2h=2\bar h$ and $a^{\Delta_1,\Delta_2}_{\Delta_3}$ is given in \eqref{eq:a}. Using \eqref{eq:It=I}, we can also obtain the spectral density $\widetilde{\rho}^{\Delta_i}_{h,\bar{h}}$ of the shadow celestial amplitudes $\widetilde{\mathcal{A}}^{\Delta_i}$ by performing shadow transformation on $\mathcal{O}_2$, $\mathcal{O}_3$ and $\mathcal{O}_4$, giving that
\begin{align}
\begin{split}\label{eq:shadow_kernel}
\frac{\widetilde{\rho}^{\Delta_i}_{h,\bar{h}}}{n_{h,\bar{h}}}K^{2-\Delta_3,2-\Delta_4}_{1-h,1-\bar{h}}=&\frac{\widetilde{\rho}^{\Delta_i}_{h,\bar{h}}}{n_{h,\bar{h}}}K^{\Delta_3,\Delta_4}_{1-h,1-\bar{h}}=-\pi \frac{e^{-i\pi\beta/2}}{\sin(\frac{-\pi\beta}{2})}C_{\Delta_1,\Delta_2,\Delta}(m)C_{\Delta_3,\Delta_4,\Delta}(m)D^{\Delta\Delta}(m)\;,
\end{split}
\end{align}
where $C_{\Delta_i,\Delta_j,\Delta}(m)$ is given in \eqref{eq:C3pt1} and we used the fact that $\beta=\sum_{i=1}^4\Delta_i-4\rightarrow -\beta=4-\sum_{i=1}^4\Delta_i$ after performing shadow transformations.\footnote{Using \eqref{eq:It=I}, one could also obtain the partial wave expansion \eqref{eq:CPE4ptA} of the celestial amplitudes $\mathcal{A}^{\Delta_i}_s$ by simply removing the factor $2^{\Delta_1}a^{\Delta_1,\Delta_2}_{\Delta}$ in \eqref{eq:FirstRho}. It is also easy to check that the partial wave expansion of $\text{Im}\,\mathcal{A}^{\Delta_i}_s$ obtained in this way agrees with the result predicted by the optical theorem \cite{Lam:2017ofc,Chang:2021wvv}.}

The poles of \eqref{eq:shadow_kernel} located at right-half $\Delta$-plane are at $\Delta=\Delta_1+\Delta_2+2n$ and $\Delta=\Delta_3+\Delta_4+2n$ with $n=0,1,2,\cdots$. This leads to the following conformal block expansion of the shadow celestial amplitude $\widetilde{\mathcal{A}}_{s}^{\Delta_i}(z_i)$
\begin{equation}\label{eq:CBE}
\widetilde{\mathcal{A}}_{s}^{\Delta_i}(z_i)=I_{12-34}\frac{e^{-i\pi\beta/2}}{2\sin(\frac{-\pi\beta}{2})}\bigg[\sum_{n=0}^{\infty}\mathscr{C}_{\Delta_1+\Delta_2+2n}^{\Delta_i}(m)G^{\Delta_i}_{\Delta_1+\Delta_2+2n}(\chi)+\mathscr{C}_{\Delta_3+\Delta_4+2n}^{\Delta_i}(m)G^{\Delta_i}_{\Delta_3+\Delta_4+2n}(\chi)\bigg]\;.
\end{equation}
Here the expanding coefficients $\mathscr{C}_{\Delta}^{\Delta_i}$ are
\begin{align}
\begin{split}
\mathscr{C}_{\Delta}^{\Delta_i}(m)=\mathcal{C}_{\Delta_1,\Delta_2,\Delta}(m)\mathcal{C}_{\Delta_3,\Delta_4,\Delta}(m)D^{\Delta\Delta}(m)\;.
\end{split}
\end{align}
As we can see from \eqref{eq:CBE}, conformal block expansion of $\widetilde{\mathcal{A}}^{\Delta_i}_s$ contains two series of scalar exchange operators with conformal dimension $\Delta=\Delta_1+\Delta_2+2n$ and $\Delta=\Delta_3+\Delta_4+2n$, respectively. It is then natural to expect that these exchange operators are double-trace operators, which take the schematic form as
\begin{align}
\mathcal{O}_i(\partial\bar{\partial})^{n}\mathcal{O}_j   
\end{align}
with conformal dimension $\Delta_i+\Delta_j+2n$. After interpreting the exchange operators in \eqref{eq:CBE} as double-trace operators, the conformal block expansion \eqref{eq:CBE} is then reminiscent of the conformal block expansion of four-point contact Witten diagrams in AdS. %The analogy between the conformal block expansion of $\widetilde{\mathcal{A}}^{\Delta_i}_s$ and four-point contact Witten diagrams is not surprised. Indeed, celestial amplitudes $\widetilde{\mathcal{A}}^{\Delta_i}$ always look like contact diagrams in the sense that internal propagators in scattering amplitudes never get mapped to be single-trace operators in CCFTs.
Moreover, from the OPE analysis in Section \ref{sec:OPE}, we already saw the double-trace operators $\mathcal{O}_1\mathcal{O}_2$ and $\mathcal{O}_3\mathcal{O}_4$, coming from the leading OPE limit $\hat{q}_1\rightarrow\hat{q}_2$ and $\hat{q}_3\rightarrow\hat{q}_4$, respectively. 
More precisely, we can recognize that the first term in \eqref{eq:CBE} recovers the result \eqref{eq:OPE12then34} from the $\hat{q}_1\rightarrow\hat{q}_2$ OPE limit which is valid when $\text{Re}(\Delta_1+\Delta_2-\Delta_3-\Delta_4)<0$.

%\fixme{Remove the following paragraph?}
%We repeat here that our OPE analysis in Section \ref{sec:OPE} the $(n-1)$-point shadow celestial amplitude is finite. In the present case, this implies that $\text{Re}(\Delta_1+\Delta_2-\Delta_3-\Delta_4)<0$ if we look at $\mathcal{O}_1\mathcal{O}_2$-OPE and $\text{Re}(\Delta_1+\Delta_2-\Delta_3-\Delta_4)>0$ if we look at $\mathcal{O}_3\mathcal{O}_4$-OPE. Thus the second term in \eqref{eq:CBE} is sub-leading when the OPE limit $\hat{q}_1\rightarrow\hat{q}_2$ is considered, leading to
%
%\begin{align}\label{eq:OPE12}
%\begin{split}
%\lim_{\hat{q}_1\rightarrow\hat{q}_2}\widetilde{\mathcal{A}}_{s}^{\Delta_i}(z_i)=&\frac{e^{-i\pi\beta/2}}{2\sin(\frac{-\pi\beta}{2})}\mathcal{C}_{\Delta_1,\Delta_2,\Delta_1+\Delta_2}(m)D^{\Delta_1+\Delta_2,\Delta_1+\Delta_2}(m)\widetilde{\mathcal{A}}^{\Delta_3,\Delta_4,\Delta_1+\Delta_2}_m\;,
%\end{split}
%\end{align}
%
%which agrees with \eqref{eq:OPES4pt}. Moreover, taking the imaginary part of \eqref{eq:OPE12} then directly reproduces our OPE analysis from the generalized optical theorem \eqref{eq:OPEO4pt}.

Although we can not see the exchange operators with conformal dimension $\Delta=\Delta_3+\Delta_4+2n$ by looking at the OPE limit $\hat{q}_1\rightarrow\hat{q}_2$ at leading order, these exchange operators can be recovered by studying the OPE limit $\hat{q}_3\rightarrow\hat{q}_4$. This makes the four-point shadow celestial amplitude $\widetilde{\mathcal{A}}^{\Delta_i}_{2\rightarrow 2}$ a bit special since we can implement our OPE analysis for both incoming and outgoing operators and recover all the exchange operators appearing in the conformal block expansion \eqref{eq:CBE}.

\section{Outlook}\label{Sec:outlook}
In this paper, we proposed the shadow conformal primary basis \eqref {eq:NewMassless} for massless scalar particles and defined the shadow celestial amplitude in \eqref{eq:NCA}. The shadow conformal primary basis can be obtained from the conformal primary basis by performing shadow transformations as in \eqref{eq:NewMassless_as_shadow}. In terms of the shadow conformal primary basis, the shadow celestial amplitude defined in \eqref{eq:NCA} behaves more like a standard CFT correlator. We studied the constraints from the translation symmetry on the shadow celestial amplitudes. Moreover, based on the factorization of the scattering amplitudes in the plane-wave basis, we further showed that the shadow celestial amplitudes enjoy a nice factorization in the leading OPE limit. Since 
%the OPE limit of the shadow celestial amplitudes is not directly performed on the momentum
the celestial coordinates in the shadow celestial amplitudes are not directly related to the momenta of the particles, the kinematic constraints like the momentum conservation do not lead to any subtlety when we study the OPE behaviour of the shadow celestial amplitude. We checked the OPE factorization by focusing on the tree-level four-point shadow celestial amplitude of four massless external scalars and one massive exchange scalar. Several interesting open questions ensue from our work. 

\begin{itemize}

\item In this paper, we only studied the OPE limit at leading order in the sense that we set $\hat{q}_1=\hat{q}_2$ directly. It would be of great interest to study the OPE limit at sub-leading orders and explore possible exchange operators. As we see, all of the exchange operators in the four-point conformal block expansion \eqref{eq:CBE} can be uncovered by looking at the leading order OPE $\hat{q}_1\rightarrow\hat{q}_2$ and $\hat{q}_3\rightarrow\hat{q}_4$. However, we expect that the leading order OPE is not enough to generate all exchange operators in higher-point amplitudes. To get the exchange operators in the higher-point conformal block expansion, one must go beyond the leading OPE.\footnote{ Recent progress on the computation of higher-point conformal blocks can be found in\cite{Alkalaev:2015fbw,Rosenhaus:2018zqn,Parikh:2019ygo,Goncalves:2019znr,Jepsen:2019svc,Parikh:2019dvm,Fortin:2019zkm,Fortin:2020yjz,Anous:2020vtw,Haehl:2021tft,Fortin:2020bfq,Hoback:2020pgj,fortin2020all,Buric:2020dyz,Hoback:2020syd,Poland:2021xjs,Buric:2021ywo,Buric:2021ttm,Buric:2021kgy,fortin2022feynman,Bercini:2020msp,Antunes:2021kmm}.}

\item The OPE analysis in Section \ref{ssec:OPEO} works only for scattering amplitude at tree-level. When the scattering amplitude $\mathcal{M}_{2\rightarrow n}$ at loop-level is considered, the intermediate states appearing in the generalized optical theorem also include multi-particle states. It would be interesting to see how to translate the factorization of the scattering amplitude in the plane-wave basis into the factorization of the corresponding shadow celestial amplitude when there exist multi-particle states. Moreover, our OPE analysis in Section \ref{ssec:OPES} is restricted to a particular class of Feynman diagrams depicted in Figure \ref{fig:Feynman}. Extending that analysis to more generic diagrams is necessary to fully understand the OPE behaviour in CCFTs. 

\item Another avenue would be to explore the flat space correspondence of the double-trace operators appearing in the conformal block expansion \eqref{eq:CBE}.  It is well-known that double-trace operators in AdS correspond to two-particle states. Although some recent progress has been made to relate celestial amplitudes with AdS Witten diagrams \cite{Casali:2022fro,Iacobacci:2022yjo}, it is unclear if this statement still holds in CCFTs. Thus it would be of great interest to study these double-trace operators and explore their holographic dual in flat space.  

\item Finally, it would be interesting to study the Mellin amplitudes associated with the shadow celestial amplitudes. Celestial Mellin amplitudes in three dimensions have been studied in \cite{Jiang:2022hho}. For the usual four-dimensional celestial amplitudes, defining the corresponding celestial Mellin amplitudes is subtle due to the existence of delta-function $\delta(\chi-\bar{\chi})$. In contrast with the celestial amplitudes, the shadow celestial amplitude defined in this paper no longer has this distributional factor and takes the standard form as a CFT correlation function. Thus, following the definition of AdS Mellin amplitudes, one can define Mellin amplitudes associated with the shadow celestial amplitudes, and the techniques developed in AdS Mellin amplitudes can be used to study the shadow celestial Mellin amplitudes.  
\end{itemize}

%\acknowledgments
\section*{Acknowledgments}
We owe our gratitude to  Jean-Fran\c cois Fortin, Juntao Wang, Fengjun Xu, and Xinan Zhou for discussions. %, and to XXX for insightful discussions and comments on the draft.
CC is partly supported by National Key R\&D Program of China (NO. 2020YFA0713000). The work of HS is supported in part by the Beijing Postdoctoral Research Foundation.
%XXX thanks XX University for its hospitality during the progression of this work.

\appendix

\section{Preliminaries}\label{sec:pre}

\subsection{The generalized optical theorem}

In quantum field theories, one of the most important observables is the $S$-matrix, which is unitary and defined as
\begin{align}
\mathcal{S}_{i\rightarrow f}\equiv\langle f|S|i\rangle
\end{align}
where we used $i$ and $f$ to label possible in- and out-states, respectively. To extract the information from the interaction, one can define the $T$-matrix by
\begin{align}
 S=1+iT\;.   
\end{align}
This leads to the definition of the scattering amplitude $\mathcal{M}_{i\rightarrow f}$ and $\mathcal{T}_{i\rightarrow f}$ 
\begin{align}
  \langle f|iT|i\rangle\equiv i\mathcal{M}_{i\rightarrow f}\equiv(2\pi)^4i\delta^{(4)}(p_f-p_i)\mathcal{T}_{i\rightarrow f}.  
\end{align}
In perturbation theory, it is the $i\mathcal{T}_{i\rightarrow f}$ that can be computed by Feynman diagrams. The unitarity of the $S$-matrix implies that
\begin{align}
i(T^{\dagger}-T)=T^{\dagger}T\;.
\end{align}
The above relation can be recast into the matrix form, giving
\begin{align}
i\langle f|T^{\dagger}-T|i\rangle=\langle f|T^{\dagger}T|i\rangle\;,
\end{align}
After inserting the completeness relation
\begin{align}
\sum_X\prod_{j\in X}\int\frac{d^3p_j}{(2\pi)^3}\frac{1}{2E_j}=1\;,
\end{align}
we get
\begin{align}
i\langle f|T^{\dagger}-T|i\rangle=\sum_X\bigg(\prod_{j\in X}\int\frac{d^3q_j}{(2\pi)^3}\frac{1}{2E_j}\bigg)\langle f|T^{\dagger}|\{q_j\}\rangle\langle\{q_j\}|T|i\rangle\;.
\end{align}
Expressing the $T$-matrix elements as the scattering amplitude $\mathcal{M}$ leads to
\begin{align}
-i(\mathcal{M}_{i\rightarrow f}-\mathcal{M}_{f\rightarrow i}^{*})=\sum_X\bigg(\prod_{j\in X}\int\frac{d^3p_j}{(2\pi)^3}\frac{1}{2E_j}\bigg)\mathcal{M}_{i\rightarrow X}\mathcal{M}^{*}_{f\rightarrow X}\;,
\end{align}
which is known as the generalized optical theorem in quantum field theory. In a unitary theory containing only real scalars, the CPT symmetry guarantees that $\mathcal{M}_{i\rightarrow f}=\mathcal{M}_{f\rightarrow i}$. Thus, the generalized optical theorem can be written as
\begin{align}\label{eq:optical}
\text{Im}\,\mathcal{M}_{i\rightarrow f}=\frac{1}{2}\sum_X\bigg(\prod_{j\in X}\int\frac{d^3p_j}{(2\pi)^3}\frac{1}{2E_j}\bigg)\mathcal{M}_{i\rightarrow X}\mathcal{M}^{*}_{X\rightarrow f}\;.
\end{align}

\subsection{Shadow transformation and conformal partial waves}\label{sec:PW}

In a $d$-dimensional conformal field theory, the shadow operator $\widetilde{\mathcal{O}}_{\widetilde{\Delta}}^{\mu_1\cdots\mu_{J}}$ with conformal dimension $\widetilde{\Delta}\equiv d-\Delta$ and spin-$J$ is defined by performing the shadow trnsformation on the conformal primary operator $\mathcal{O}_{\Delta}^{\mu_1\cdots\mu_J}$, \textit{i.e.},
\begin{align}
\widetilde{\mathcal{O}}_{d-\Delta}^{\mu_1\cdots\mu_{J}}(x)=k_{\Delta,J}\int d^dy\frac{\mathcal{I}^{\mu_1\cdots\mu_J,\nu_1\cdots\nu_J}(x-y)}{(x-y)^{2(d-\Delta)}}\mathcal{O}_{\Delta,\nu_1\cdots\nu_J}(y)\;,
\end{align}
where $\mathcal{I}^{\mu_1\cdots\mu_J,\nu_1\cdots\nu_J}(x)$ is the symmetric traceless inversion tensor that can be constructed from the products of $I^{\mu\nu}(x)=g^{\mu\nu}-2x^{\mu}x^{\nu}/x^2$, and $k_{\Delta,J}$ is
\begin{align}
k_{\Delta,J}=\frac{\Gamma[d-\Delta+J]}{\pi^{\frac{d}{2}}(\Delta-1)_J\Gamma[\Delta-\frac{d}{2}]}\;.
\end{align}
Armed with shadow operators, the following resolution of the identity in Euclidean CFTs can be obtained by the harmonic analysis of the conformal group $SO(d+1,1)$
\begin{align}\label{eq:ROI}
|0\rangle\langle0|+\sum_{J}\int_{\frac{d}{2}}^{\frac{d}{2}+i\infty}\frac{d\Delta}{2\pi i}\int d^dx\mathcal{O}_{\Delta}^{\mu_1\cdots\mu_{J}}(x)|0\rangle\langle0|\widetilde{\mathcal{O}}_{d-\Delta,\mu_1\cdots\mu_{J}}(x)+\cdots=1\;,
\end{align}
where $\cdots$ represents complementary and discrete series. Inserting the resolution of the identity into a scalar four-point function in two dimensional Euclidean CFTs then leads to the conformal partial wave expansions,\footnote{$\mathcal{O}_i(z_i)$ should be understood as $\mathcal{O}_i(z_i,\bar{z}_i)$. We use this abbreviation to simplify the notation.}
\begin{align}\label{eq:CPWE}
\begin{split}
\langle\mathcal{O}_1(z_1)\cdots\mathcal{O}_4(z_4)\rangle=&\sum_{J=-\infty}^{+\infty}\int_{0}^{+\infty}\frac{d\lambda}{2\pi}\frac{\rho^{\Delta_i}_{h,\bar{h}}}{n_{h,\bar{h}}}\Psi^{\Delta_i}_{h,\bar{h}}(z_i)+\cdots\;,\hspace{0.5cm}n_{h,\bar{h}}=\frac{-2\pi^3}{(2h-1)(2\bar{h}-1)}\;,
\end{split}
\end{align}
where $h\equiv(\Delta+J)/2=(1+i\lambda+J)/2$, $\bar{h}\equiv(\Delta-J)/2=(1+i\lambda-J)/2$ and the conformal partial wave $\Psi^{\Delta_i}_{h,\bar{h}}(z_i)$ is defined as
\begin{align}\label{eq:IR}
\Psi^{\Delta_i}_{h,\bar{h}}(z_i)=\int d^2z_0\langle\mathcal{O}_1(z_1)\mathcal{O}_2(z_2)\mathcal{O}_{h,\bar{h}}(z_0)\rangle\langle\tilde{\mathcal{O}}_{1-h,1-\bar{h}}(z_0)\mathcal{O}_3(z_3)\mathcal{O}_{4}(z_4)\rangle\;.
\end{align}
Evaluating the integral in \eqref{eq:IR} gives an explicit expression for conformal partial waves,
\begin{align}\label{eq:Psi=G}
\begin{split}
&\Psi^{\Delta_i}_{h,\bar{h}}(z_i)=I_{12-34}(z_i)\bigg(K^{\Delta_3,\Delta_4}_{1-h,1-\bar{h}}G^{\Delta_i}_{h,\bar{h}}(\chi)+K^{\Delta_1,\Delta_2}_{h,\bar{h}}G^{\Delta_i}_{1-h,1-\bar{h}}(\chi)\bigg)\;,
\end{split}
\end{align}
where $G^{\Delta_i}_{h,\bar{h}}(\chi)$ are the conformal blocks, which are functions of conformal cross-ratios
\begin{align}
\chi=\frac{z_{12}z_{34}}{z_{13}z_{24}}\;,\hspace{0.5cm}\bar{\chi}=\frac{\bar{z}_{12}\bar{z}_{34}}{\bar{z}_{13}\bar{z}_{24}}\;,
\end{align}
and the coefficients $K$ are given by
\begin{align}
&K^{\Delta_1,\Delta_2}_{h,\bar{h}}=(-1)^{J}\frac{\pi\Gamma[2h-1]\Gamma[1-\bar{h}+h_{12}]\Gamma[1-\bar{h}-h_{12}]}{\Gamma[2-2\bar{h}]\Gamma[h+h_{12}]\Gamma[h-h_{12}]}\;.
\end{align}
$I_{12-34}(z_i)$ in \eqref{eq:Psi=G} is
\begin{align}\label{eq:I1234}
  I_{12-34}\equiv\dfrac{\left(\dfrac{z_{24}}{z_{14}} \right)^{h_{12}} \left(\dfrac{z_{14}}{z_{13}} \right)^{h_{34}}}{z_{12}^{h_1 + h_2} z_{34}^{h_3+h_4}}\dfrac{\left(\dfrac{\bar{z}_{24}}{\bar{z}_{14}} \right)^{\bar{h}_{12}} \left(\dfrac{\bar{z}_{14}}{\bar{z}_{13}} \right)^{\bar{h}_{34}}}{\bar{z}_{12}^{\bar{h}_1+\bar{h}_2} \bar{z}_{34}^{\bar{h}_3+\bar{h}_4}}\;,\quad h_{ij}\equiv h_i-h_j\;,\quad\bar{h}_{ij}\equiv \bar{h}_i-\bar{h}_j\;.
\end{align}
Plugging \eqref{eq:Psi=G} into \eqref{eq:CPWE} and using the following property of the spectral density $\rho^{\Delta_i}_{h,\bar{h}}$
\begin{align}\label{eq:I=I}
\rho_{h,\bar{h}}^{\Delta_i}K^{\tilde{\Delta}_3,\tilde{\Delta}_4}_{1-h,1-\bar{h}}=\rho_{1-h,1-\bar{h}}^{\Delta_i}K^{\tilde{\Delta}_1,\tilde{\Delta}_2}_{1-h,1-\bar{h}}\;,
\end{align}
we get 
\begin{align}\label{eq:CPWE1}
\begin{split}
\langle\mathcal{O}_1(z_1)\cdots\mathcal{O}_4(z_4)\rangle=&\sum_{J=-\infty}^{+\infty}\int_{-\infty}^{+\infty}\frac{d\lambda}{2\pi}\frac{\rho^{\Delta_i}_{h,\bar{h}}}{n_{h,\bar{h}}}K^{\Delta_3,\Delta_4}_{1-h,1-\bar{h}}G^{\Delta_i}_{h,\bar{h}}(z_i)
+\cdots\;.
\end{split}
\end{align}
The conformal block expansion then can be obtained by enclosing the contour to the right-half $\lambda$-plane in \eqref{eq:CPWE1}. After defining 
\begin{align}
&\langle\mathcal{O}_1(z_1)\cdots\mathcal{O}_4(z_4)\rangle=I_{12-34}f(\chi)\;,\hspace{1.5cm}\Psi^{\Delta_i}_{h,\bar{h}}(z_i)=I_{12-34}\Psi^{\Delta_i}_{h,\bar{h}}(\chi)\;,
\end{align}
the conformal partial wave expansion then can be written as
\begin{align}
f(\chi)=\sum_{J=-\infty}^{+\infty}\int_{0}^{+\infty}\frac{d\lambda}{2\pi}\frac{\rho^{\Delta_i}_{h,\bar{h}}}{n_{h,\bar{h}}}\Psi^{\Delta_i}_{h,\bar{h}}(\chi)+\cdots\;,
\end{align}

The conformal partial wave expansion for four-point functions involving a shadow operator can be easily derived from \eqref{eq:CPWE} and \eqref{eq:IR} when there are only scalar ($J=0$) partial waves. Specifically, starting with \eqref{eq:CPWE} and performing shadow transformation for the first operator $\mathcal{O}_1(z_1)$ leads to
\begin{align}\label{eq:SCPWE}
\begin{split}
\langle\widetilde{\mathcal{O}}_1(z_1)\cdots\mathcal{O}_4(z_4)\rangle=&\int d^2z_{1^{\prime}}\frac{1}{(z_{11^{\prime}}^2)^{\widetilde{\Delta}_1}}\int_{0}^{+\infty}\frac{d\lambda}{2\pi}\frac{\rho^{\Delta_i}_{h,\bar{h}}}{n_{h,\bar{h}}}\Psi^{\Delta_i}_{h,\bar{h}}(z_{1^{\prime}},z_2,z_3,z_4)+\cdots\;.
\end{split}
\end{align}
Using the integral representation \eqref{eq:IR} of the conformal partial waves, we find that
\begin{align}
\begin{split}
&\int d^2z_{1^{\prime}}\frac{1}{(z_{11^{\prime}}^2)^{\widetilde{\Delta}_1}}\Psi^{\Delta_i}_{h,\bar{h}}(z_{1^{\prime}},z_2,z_3,z_4)\\
=&\int d^2z_{1^{\prime}}\frac{1}{(z_{11^{\prime}}^2)^{\widetilde{\Delta}_1}}\int d^2z_0\frac{1}{(z^2_{1^{\prime}2})^{\frac{\Delta_1+\Delta_2-\Delta}{2}}(z^2_{01^{\prime}})^{\frac{\Delta+\Delta_1-\Delta_2}{2}}(z^2_{02})^{\frac{\Delta+\Delta_2-\Delta_1}{2}}}\\
&\qquad\times\frac{1}{(z^2_{03})^{\frac{\widetilde{\Delta}+\Delta_3-\Delta_4}{2}}(z^2_{04})^{\frac{\widetilde{\Delta}+\Delta_4-\Delta_3}{2}}(z^2_{34})^{\frac{\Delta_3+\Delta_4-\widetilde{\Delta}}{2}}}\;.
\end{split}
\end{align}
The integral over $z_{1^{\prime}}$ can be computed by noting that
\begin{align}\label{eq:identity1}
\begin{split}
&\int d^2z_{1^{\prime}}\frac{1}{(z_{11^{\prime}}^2)^{\widetilde{\Delta}_1}(z^2_{1^{\prime}2})^{\frac{\Delta_1+\Delta_2-\Delta}{2}}(z^2_{01^{\prime}})^{\frac{\Delta+\Delta_1-\Delta_2}{2}}(z^2_{02})^{\frac{\Delta+\Delta_2-\Delta_1}{2}}}\\
=&a^{\Delta_1,\Delta_2}_{\Delta}\frac{1}{(z^2_{12})^{\frac{\widetilde{\Delta}_1+\Delta_2-\Delta}{2}}(z^2_{01})^{\frac{\Delta+\widetilde{\Delta}_1-\Delta_2}{2}}(z^2_{02})^{\frac{\Delta+\Delta_2-\widetilde{\Delta}_1}{2}}}\;,
\end{split}
\end{align}
where $a^{\Delta_1,\Delta_2}_{\Delta}$ is
\begin{align}\label{eq:a}
a^{\Delta_1,\Delta_2}_{\Delta}=\frac{\pi\Gamma[\Delta_1-1]\Gamma[\frac{\Delta+\widetilde{\Delta}_1-\Delta_2}{2}]\Gamma[\frac{\widetilde{\Delta}_1+\Delta_2-\Delta}{2}]}{\Gamma[2-\Delta_1]\Gamma[\frac{\Delta_1+\Delta_2-\Delta}{2}]\Gamma[\frac{\Delta+\Delta_1-\Delta_2}{2}]}\;.
\end{align}
This leads to the following conformal partial wave expansion for $\langle\widetilde{\mathcal{O}}_1(z_1)\cdots\mathcal{O}_4(z_4)\rangle$
\begin{align}
\begin{split}
\langle\widetilde{\mathcal{O}}_1(z_1)\cdots\mathcal{O}_4(z_4)\rangle=&\int_{0}^{+\infty}\frac{d\lambda}{2\pi}\frac{\rho^{\widetilde{\Delta}_1,\Delta_2,\Delta_3,\Delta_4}_{h,\bar{h}}}{n_{h,\bar{h}}}\Psi^{\widetilde{\Delta}_1,\Delta_2,\Delta_3,\Delta_4}_{h,\bar{h}}(z_i)+\cdots\;,
\end{split}
\end{align}
where the new spectral density $\rho^{\widetilde{\Delta}_1,\Delta_2,\Delta_3,\Delta_4}_{h,\bar{h}}$ is related to the old spectral density $\rho^{\Delta_1,\Delta_2,\Delta_3,\Delta_4}_{h,\bar{h}}$ through
\begin{align}\label{eq:It=I}
\begin{split}
\rho^{\widetilde{\Delta}_1,\Delta_2,\Delta_3,\Delta_4}_{h,\bar{h}}=&a^{\Delta_1,\Delta_2}_{\Delta}\rho^{\Delta_1,\Delta_2,\Delta_3,\Delta_4}_{h,\bar{h}}\;.
\end{split}
\end{align}
%
%%%%%%%%%%%%%%%%%%%%%%%%%%%%%%%%%%%%%%%%%%%%%%%%%%%%%%%%%%%%%%%%%%%%%%%%%%%

\section{The conformal integral}\label{sec:ConformalIntegral}

In this appendix, we will compute the following conformal integral
\begin{align}
I(\hat{q}_1,\hat{q}_2,\hat{p})\equiv\int D^2\hat{q}^{\prime}_2\frac{1}{(-\hat{q}_2\cdot\hat{q}^{\prime}_2)^{\Delta_2}(-\hat{q}^{\prime}_2\cdot\hat{p})^{2-\Delta_1-\Delta_2}(-\hat{q}^{\prime}_2\cdot Y)^{\Delta_1}}\;,
\end{align}
where $D^2\hat{q}^{\prime}=d^2z^{\prime}$, $Y^{\mu}=-2(\hat{q}_{1}\cdot\hat{p})\hat{p}^{\mu}-\hat{q}_{1}^{\mu}$. Using the Feynman/Schwinger parameterization, $I(\hat{q}_1,\hat{q}_2,\hat{p})$ can be written as
\begin{align}\label{eq:1}
I(\hat{q}_1,\hat{q}_2,\hat{p})=\frac{1}{\Gamma[\Delta_1]\Gamma[\Delta_2]\Gamma[2-\Delta_1-\Delta_2]}\int_{0}^{\infty}dx_1dx_2\int D^2\hat{q}^{\prime}_2\frac{x_1^{\Delta_1-1}x_2^{\Delta_2-1}}{[-\hat{q}^{\prime}_2\cdot(\hat{p}+x_1 Y+x_2\hat{q}_2)]^2}\;.
\end{align}
We evaluate the integral over $\hat{q}_2^{\prime}$ by noting that \cite{Simmons-Duffin:2012juh}
\begin{align}\label{eq:2}
   \int D^2\hat{q}^{\prime}_2\frac{1}{(-\hat{q}^{\prime}_2\cdot Q)^2}=\frac{2\pi}{-Q^2}\;,
\end{align}
which holds for any timelike vector $Q^{\mu}$. This can be seen by going to the rest frame of $Q^{\mu}$, \textit{i.e.} we choose $Q^{\mu}=(1,0,0,0)$. In this frame, the above integral becomes
\begin{align}
   \int D^2\hat{q}^{\prime}_2\frac{1}{(-\hat{q}^{\prime}_2\cdot Q)^2}=\int d^2z\frac{1}{(1+z\bar{z})^2}=2\pi\;.
\end{align}
Recovering the dependence on $Q^2$ by dimensional analysis then gives \eqref{eq:2}. Plugging \eqref{eq:2} back into \eqref{eq:1} leads to
\begin{align}
\begin{split}
I(\hat{q}_1,\hat{q}_2,\hat{p})=&\frac{2\pi}{\Gamma[\Delta_1]\Gamma[\Delta_2]\Gamma[2-\Delta_1-\Delta_2]}\\
&\times\int_{0}^{\infty}\frac{dx_1dx_2\,x_1^{\Delta_1-1}x_2^{\Delta_2-1}}{(1-2x_1\hat{q}_1\cdot\hat{p}-2x_2\hat{q}_2\cdot\hat{p}+4x_1x_2\hat{q}_1\cdot\hat{p}\hat{q}_2\cdot\hat{p}-4x_1x_2\hat{q}_{12})}\;.
\end{split}
\end{align}
Evaluating the integral over $x_1$, $x_2$ and expanding the results around small $\hat{q}_{12}$ finally lead to
\begin{align}\label{eq:ConformalI}
\begin{split}
I(\hat{q}_1,\hat{q}_2,\hat{p})=\sum_{n=0}^{\infty}\frac{2\pi\Gamma[1-\Delta_1]\Gamma[1-\Delta_2](\Delta_1)_n(\Delta_2)_n}{2^{\Delta_1+\Delta_2}\Gamma[2-\Delta_1-\Delta_2]\Gamma[n+1]\Gamma[n+1]}\frac{(-\hat{q}_{12})^n}{(\hat{q}_1\cdot\hat{p})^{\Delta_1+n}(-\hat{q}_2\cdot\hat{p})^{\Delta_2+n}}\;.
\end{split}
\end{align}
Setting $\hat{q}_{1}=\hat{q}_2$ in \eqref{eq:ConformalI} leads to
\begin{align}\label{eq:ConformalI1}
\begin{split}
I(\hat{q}_2,\hat{q}_2,\hat{p})=\frac{2\pi\Gamma[1-\Delta_1]\Gamma[1-\Delta_2]}{2^{\Delta_1+\Delta_2}\Gamma[2-\Delta_1-\Delta_2]}\;.
\end{split}
\end{align}
%


\bibliographystyle{ssg}
\bibliography{Shadow}

% \bibitem{Bondi:1962px}
%   H.~Bondi, M.~G.~J.~van der Burg and A.~W.~K.~Metzner,
%   ``Gravitational waves in general relativity. 7. Waves from axisymmetric isolated systems,''
%   Proc.\ Roy.\ Soc.\ Lond.\ A {\bf 269}, 21 (1962).
  
% \bibitem{Sachs:1962wk}
%   R.~K.~Sachs,
%   ``Gravitational waves in general relativity. 8. Waves in asymptotically flat space-times,''
%   Proc.\ Roy.\ Soc.\ Lond.\ A {\bf 270}, 103 (1962).
  
% \bibitem{Barnich:2009se}
%   G.~Barnich and C.~Troessaert,
%   ``Symmetries of asymptotically flat 4 dimensional spacetimes at null infinity revisited,''
%   Phys.\ Rev.\ Lett.\  {\bf 105}, 111103 (2010)
%   %doi:10.1103/PhysRevLett.105.111103
%   [arXiv:0909.2617 [gr-qc]].

\end{document}